\documentclass[preprint,12pt,3p,authoryear]{elsarticle}

\usepackage{amsfonts}
\usepackage{bbm}
\usepackage{dsfont}
\usepackage{amssymb}
\usepackage{amsmath}
\usepackage{graphicx}
\usepackage[retainorgcmds]{IEEEtrantools}
\usepackage[colorlinks]{hyperref}
\hypersetup{
  citecolor = {blue}
}
\usepackage{siunitx}
\usepackage{pdflscape}

\usepackage{enumitem}
\usepackage[normalem]{ulem}
\useunder{\uline}{\ul}{}
\usepackage{setspace} 
\usepackage[utf8]{inputenc}
\usepackage[linesnumbered, ruled, vlined]{algorithm2e}
\usepackage{nicefrac}
\usepackage{xspace}
\usepackage{float}
\usepackage{caption}
\usepackage{subcaption}
\usepackage{rotating, lscape, longtable, tabu, booktabs, siunitx, multirow}
\usepackage[capitalise, noabbrev]{cleveref}
\usepackage{verbatim}
\usepackage{mathtools}
\usepackage{placeins}

\usepackage{tkz-graph}
\usetikzlibrary{positioning,calc}

\usetikzlibrary{arrows,arrows.meta,shapes,decorations.pathmorphing, decorations.pathreplacing, decorations.markings, positioning}
\let\svtikzpicture\tikzpicture
\def\tikzpicture{\noindent\svtikzpicture}

\newcommand{\uset}[1]{\ifmmode\left\{\,#1\,\right\}\else\{\,#1\,\}\fi}
\newcommand{\ulst}[1]{\ifmmode\left[\,#1\,\right]\else[\,#1\,]\fi}
\newcommand{\upar}[1]{\ifmmode\left(\,#1\,\right)\else(\,#1\,)\fi}
\newcommand{\uioc}[1]{\ifmmode\left(\,#1\,\right]\else(\,#1\,]\fi}
\newcommand{\uico}[1]{\ifmmode\left[\,#1\,\right)\else[\,#1\,)\fi}

\newtheorem{theorem}{Theorem}
\newtheorem{lemma}[theorem]{Lemma}

\newenvironment{proof}{\noindent\textbf{Proof}.}{\hfill$\square$}

\newtheorem{definition}{Definition}
\newtheorem{example}{Example}
\newtheorem{remark}{Remark}
\newtheorem{problem}{Problem}

\crefname{theorem}{Theorem}{Theorems}
\Crefname{theorem}{Theorem}{Theorems}

\crefname{lemma}{Lemma}{Lemmas}
\Crefname{lemma}{Lemma}{Lemmas}

\crefname{proposition}{Proposition}{Propositions}
\Crefname{proposition}{Proposition}{Propositions}

\crefname{corollary}{Corollary}{Corollaries}
\Crefname{corollary}{Corollary}{Corollaries}

\crefname{conjecture}{Conjecture}{Conjectures}
\Crefname{conjecture}{Conjecture}{Conjectures}

\crefname{definition}{Definition}{Definitions}
\Crefname{definition}{Definition}{Definitions}

\crefname{example}{Example}{Examples}
\Crefname{example}{Example}{Examples}

\crefname{remark}{Remark}{Remarks}
\Crefname{remark}{Remark}{Remarks}

\crefname{problem}{Problem}{Problems}
\Crefname{problem}{Problem}{Problems}

\journal{Transportation Research Part C: Emerging Technologies}

\begin{document}

\begin{frontmatter}

\title{Vehicle Rebalancing Under Adherence Uncertainty}

\author[aff1]{Avalpreet Singh Brar\corref{cor1}}
\ead{brar0002@e.ntu.edu.sg}

\author[aff1]{Rong Su}
\ead{rsu@ntu.edu.sg}

\author[aff2]{Christos G. Cassandras}
\ead{cgc@bu.edu}

\author[aff3]{Max Ng}
\ead{manng.tsz@gmail.com}

\author[aff4]{\texorpdfstring{\\}{ }Yuling Li}
\ead{yuling@ustb.edu.cn}

\author[aff5]{Gioele Zardini}
\ead{gzardini@mit.edu}

\cortext[cor1]{Corresponding author}

\address[aff1]{School of Electrical and Electronic Engineering, 
Nanyang Technological University, Singapore}

\address[aff2]{Division of Systems Engineering, Boston University, USA}

\address[aff3]{Uber Technologies, Inc., San Francisco, CA, USA}

\address[aff4]{School of Automation and Electrical Engineering, 
University of Science and Technology Beijing, China}

\address[aff5]{Laboratory for Information and Decision Systems, 
Massachusetts Institute of Technology, USA}

\tnotetext[t1]{This study is supported under the RIE2020 Industry Alignment Fund – Industry Collaboration Projects (IAF-ICP) Funding Initiative, as well as cash and in-kind contribution from the industry partner(s).}

\tnotetext[t2]{Rong Su's research is supported by the National Research Foundation Singapore under its AI Singapore Programme (Award Number: AISG2-GC-2023-007) and the National Research Foundation, Singapore through its Medium Sized Center for Advanced Robotics Technology Innovation (CARTIN) under Project WP2.7.}

\begin{abstract}
Ride-hailing platforms frequently face spatiotemporal supply-demand imbalances caused by both uneven passenger demand across locations and decentralized, uncoordinated driver decision-making. 
To mitigate these inefficiencies, existing fleet rebalancing methods provide repositioning recommendations to idle drivers, but most assume that drivers always follow the assigned recommendations. 
Some recent studies relax this assumption by incorporating driver destination preferences or modeling adherence using static adherence probabilities. 
In reality, driver adherence evolves dynamically through repeated interactions with the recommender system. Existing adherence-aware methods typically generate recommendations sequentially by updating recommendations after observing each driver's response and do not jointly capture driver preferences, dynamically evolving adherence, and simultaneous fleet-wide recommendation generation.
We address this gap through the proposed Adherence-Aware Vehicle Rebalancing (AAVR) model, which generates simultaneous fleet-wide repositioning recommendations while explicitly accounting for driver preferences and dynamically evolving adherence. This formulation gives rise to a computationally intractable optimization problem. We derive a tractable reformulation based on an upper-bound relaxation, enabling real-time recommendation generation for large-scale ride-hailing systems.
Extensive simulations on the NYC taxi dataset under dynamic adherence updates demonstrate that AAVR consistently outperforms state-of-the-art vehicle rebalancing strategies, improving served demand by 26.72\%, reducing passenger waiting time by 26.45\%, increasing platform and driver profits by 25.90\% and 28.75\%, respectively, and improving the fleet's adherence probability by 30.06\%.
These results demonstrate that explicitly accounting for evolving driver adherence improves both immediate operational performance and longer-term adherence to platform recommendations.
\end{abstract}


\end{frontmatter}


\section{Introduction}
\noindent Urban mobility systems face persistent congestion and inefficient use of road resources, particularly as travel demand becomes increasingly concentrated across space and time. Ride-hailing services offer a flexible Mobility-on-Demand (MoD) solution, yet their performance is fundamentally constrained by spatiotemporal supply–demand imbalances~\citep{zardini2022analysis}. 
Vehicle rebalancing aims to address this issue by repositioning idle drivers to locations with anticipated demand shortages.
Most existing rebalancing models assume full driver compliance with platform recommendations. In practice, however, drivers may accept or reject these suggestions based on individual destination preferences and their trust in the recommender system. Moreover, such trust is not static but evolves over time in response to past experiences. To capture these behavioral dynamics, we propose an \emph{Adherence-Aware Vehicle Rebalancing (AAVR)} framework that incorporates driver-level adherence decisions and explicitly accounts for uncertainty in realized supply.

\subsection{Literature Review}
Vehicle rebalancing has been widely studied as a core mechanism for mitigating spatial supply--demand imbalance in mobility-on-demand systems. 
Early work focused primarily on aggregate flow-control formulations. \cite{pavone2012robotic} developed a station-based dispatching framework to stabilize customer queues, while \cite{spieser2016shared} incorporated both waiting customers and future requests into shared-mobility rebalancing. 
\cite{smith2013rebalancing} extended this direction by jointly routing vehicles and rebalancing drivers. 
Later, \cite{miao2017data} introduced a multi-period stochastic rebalancing model to account for the downstream effects of decisions. 
Agent-level formulations were subsequently developed to generate taxi-specific rebalancing decisions, including \cite{miao2015taxi, wallar2018vehicle}. 
These studies established the importance of proactive supply relocation, but they largely treat rebalancing decisions as commands implemented by the fleet, thereby abstracting away the behavioral response of individual drivers.

Recent work has emphasized robust and learning-based rebalancing under uncertainty. 
\cite{brar2021dynamic} adaptively selected operational parameters such as rebalancing frequency and planning horizon under uncertain forecasts. 
\citep{wen2017rebalancing, qian2022drop} applied reinforcement learning to repositioning decisions, \cite{levin2022general} characterized maximum-stability dispatch policies for shared autonomous vehicle systems and \cite{li2025learning} proposed a joint rebalancing and dynamic pricing policies for autonomous mobility-on-demand. 
\cite{kim2024estimate} connected to demand-estimation studies that emphasize exogenous factors and uncertainty quantification in mobility demand prediction. 
\cite{zhu2024coverage} formulated idle-vehicle rebalancing as a coverage-control problem, while \cite{lee2025assessing} examined the role of ride-hailing rebalancing in improving multi-modal transportation-system resilience under disruptions. 

A related body of work extends rebalancing beyond conventional taxi systems to car-sharing, electric mobility, shared micro-mobility, and broader platform-control settings. In car-sharing systems, rebalancing has been coupled with staff relocation~\cite{nourinejad2015vehicle}; in electric mobility, with charging and battery-related operations~\citep{guo2020vehicle,guo2023vehicle,brar2022supply,yu2025optimal,bogyrbayeva2021reinforcement}; and in shared micro-mobility, with rider crowd-sourcing and third-party logistics~\citep{jin2023vehicle}. 
Other studies jointly consider pricing, fleet sizing, user access trips, personnel dispatch, and real-time relocation under uncertainty~\citep{huang2021innovative,li2025real}. 
A parallel stream embeds rebalancing within platform-control models, including macroscopic fluid and market-equilibrium models~\citep{xu2021generalized,wang2021aggregate}, pricing mechanisms~\citep{li2021spatial,wollenstein2020joint,brar2024integrated}, matching or dispatch-integrated formulations~\citep{guo2021robust,feng2022approximating,azadeh2022choice,tuncel2023integrated,huang2025data, brar2025maximal}, and routing \citep{alonso2017predictive, 11423297}. 
Together, these studies address operational uncertainty and platform interactions, but not the supply-side uncertainty induced by heterogeneous driver adherence to repositioning recommendations.

Human factors are central to recommendation-based rebalancing. Several studies have analyzed taxi and ride-sourcing drivers' search behavior, operating strategies, and location preferences. Vacant taxi drivers' destination choices and taxi-stand choices have been modeled using discrete-choice models \citep{wong2014bi,wong2015two,demissie2020modeling}. GPS traces have also been used to characterize taxi service strategies, next-destination choices, and revenue-oriented operating patterns \citep{zhang2014understanding,rossi2019modelling,liu2021mining,millard2023ridehail}. Driver experience and preference evolution have been studied using data-driven and learning-based methods: taxi drivers' learning curves and preference dynamics have been analyzed from operational data, while imitation learning has been used to infer location-dependent passenger-seeking strategies \citep{pan2019dissecting,pan2020dhpa,zhang2019unveiling,zhang2020cgail}. Complementary qualitative evidence from ride-sourcing drivers further shows that acceptance, relocation, working area, and working schedule are shaped by platform policies, driver characteristics, rider attributes, and external conditions \citep{ashkrof2020understanding}. These studies demonstrate that driver repositioning behavior is structured rather than random. However, most of them model drivers' self-directed behavior or preferences, rather than how such preferences interact with platform-generated repositioning recommendations.

A closer line of work develops recommendation, dispatching, and matching methods that account for driver utility or preferences. GPS-data-driven recommendation and route-discovery methods have been proposed to help drivers find passengers or reliable routes \citep{qu2014cost,xu2014taxi,he2017collaborative}. Preference-aware taxi dispatch has been formulated as an online stable matching problem, and alternative platform designs have been compared in settings where drivers either receive information or are directly assigned recommendations, explicitly recognizing drivers' destination preferences \citep{zhao2019preference,sun2020taxi}. Recent studies have further incorporated preferences into dispatching, repositioning, and rebalancing decisions. These include models for balancing driver and passenger preferences in real-time matching, preference-aware vehicle repositioning recommendations, human preference uncertainty in shared-micromobility rebalancing and charging, and driver-behavior-aware reinforcement learning for vehicle dispatching \citep{yang2024future,zhou2024preference,tan2024human,han2025garlic}. Most closely related to our setting, taxi repositioning with probabilistic driver compliance and survey-based driver-preference-aware repositioning recommendation models have been proposed \citep{lindstroem2025taxi,chen2024rebalance}.

These studies move toward human-aware recommendation design, but two important limitations remain. First, adherence is often treated as static, even though a driver's willingness to follow recommendations evolves through repeated interactions with the platform. In practice, adherence is shaped not only by destination preferences but also by the driver's past experience with the recommender system. Such behavioral adaptation has been widely modeled using Beta-Bayesian frameworks in automation and recommender systems \citep{josang2002beta,guo2021modeling,guo2021reverse,rodriguez2023review}, but has not been incorporated into taxi rebalancing optimization, where driver adherence directly determines whether the recommended repositioning of supply is actually realized. Second, recommendation-based rebalancing models \citep{zhou2024preference,chen2024rebalance} typically generate recommendations sequentially by updating recommendations after observing each driver's response. This can be limiting in practice because drivers may not respond immediately, making their decisions unavailable when subsequent recommendations are generated. In contrast, we consider a centralized fleet-level recommendation framework that simultaneously computes repositioning recommendations for all available drivers within each planning epoch, allowing adherence uncertainty to be incorporated before any individual driver responses are observed. Together, these limitations imply that recommendation-based rebalancing is a behavioral-stochastic supply problem: recommendations do not deterministically create supply at the target region, but instead influence the probability that each driver repositions there. This paper addresses this gap by formulating vehicle rebalancing as an adherence-aware optimization problem that jointly models driver preferences and dynamic adherence, enabling more effective rebalancing recommendations.

The main contributions are summarized as follows:
\begin{enumerate}
    \item We develop a driver adherence model in which the probability of following a rebalancing recommendation depends jointly on destination preference and dynamically evolving confidence in the recommender system. To the best of our knowledge, this is the first vehicle rebalancing framework to explicitly model recommendation adherence through a dynamic confidence mechanism that evolves based on past recommendation outcomes.
    
    \item We formulate the \emph{Adherence-Aware Vehicle Rebalancing (AAVR)} problem, which captures supply-side uncertainty induced by stochastic driver adherence and dynamically evolving driver confidence. To address the resulting non-convex and computationally intractable optimization problem, we derive a tractable formulation using an upper-bound relaxation, enabling implementation in large-scale ride-hailing systems.
    
    \item We evaluate the proposed framework on a simulated taxi network calibrated with New York City taxi data, showing improved allocation efficiency and system performance over state-of-the-art rebalancing baselines.
\end{enumerate}

The remainder of this paper is organized as follows. \Cref{sec:math_form} presents the mathematical formulation of the proposed \emph{Adherence-Aware Vehicle Rebalancing (AAVR)} framework. \Cref{sec:theoretical} develops the theoretical foundations of the AAVR problem and derives a decomposition-based solution approach that enables tractable real-time implementation. Experimental results and performance evaluations are presented in \Cref{sec:results}. Finally, \Cref{sec:conc} concludes the paper and discusses directions for future research.

\section{Mathematical Formulation of Adherence Aware Vehicle Rebalancing (AAVR)}
\label{sec:math_form}
The proposed Adherence-Aware Vehicle Rebalancing (AAVR) framework is designed as a closed-loop recommendation system for idle driver repositioning. Unlike conventional rebalancing models that primarily rely on demand and travel time estimates, AAVR also accounts for driver-side behavioral factors, including destination preference and confidence in the recommender system. These components are integrated to generate personalized repositioning recommendations. The observed driver responses and allocation outcomes are then used as feedback to update driver confidence over time. \Cref{fig:aavr_architecture} provides a high-level overview of this architecture.

\begin{figure}[tbh]
\centering
\resizebox{\columnwidth}{!}{%
\begin{tikzpicture}[
    block/.style={
        draw,
        rounded corners=4pt,
        align=center,
        minimum width=3.3cm,
        minimum height=0.95cm,
        fill=gray!8,
        thick
    },
    core/.style={
        draw,
        rounded corners=6pt,
        align=center,
        minimum width=4.8cm,
        minimum height=1.15cm,
        fill=blue!8,
        thick
    },
    output/.style={
        draw,
        rounded corners=4pt,
        align=center,
        minimum width=4.2cm,
        minimum height=0.95cm,
        fill=green!8,
        thick
    },
    feedback/.style={
        draw,
        rounded corners=4pt,
        align=center,
        minimum width=4.2cm,
        minimum height=0.95cm,
        fill=orange!10,
        thick
    },
    arrow/.style={->, thick, >=Stealth},
    fb/.style={->, thick, dashed, >=Stealth},
    node distance=1.4cm and 1.5cm
]

\node[block] (demand) {Demand Prediction\\Model};
\node[block, right=of demand] (travel) {Travel Time\\Prediction Model};
\node[block, right=of travel] (pref) {Driver Preference\\Model};
\node[block, right=of pref] (conf) {Driver Confidence\\Model};

\coordinate (center) at ($(travel)!0.5!(pref)$);

\node[core, below=2.0cm of center] (aavr) {Adherence-Aware Vehicle\\Rebalancing Model};

\node[output, below=1.4cm of aavr] (rec) {Personalized Repositioning\\Recommendations};

\node[feedback, below=1.4cm of rec] (state) {Observed Driver Response\\and Allocation Outcome};

\draw[arrow] (demand.south) -- (aavr.north);
\draw[arrow] (travel.south) -- (aavr.north);
\draw[arrow] (pref.south) -- (aavr.north);
\draw[arrow] (conf.south) -- (aavr.north);

\draw[arrow] (aavr.south) -- (rec.north);
\draw[arrow] (rec.south) -- (state.north);

\draw[fb] (state.east) -| (conf.south);

\end{tikzpicture}%
}
\caption{Architecture of the proposed AAVR framework with state feedback.}
\label{fig:aavr_architecture}
\end{figure}

\begin{figure}[tb]
\centering
\resizebox{\columnwidth}{!}{%
\begin{tikzpicture}[
  xscale=1.05, yscale=1.0,
  box/.style={draw, rectangle, minimum width=0.25cm, minimum height=0.35cm, inner sep=0.5pt, anchor=center, fill=black},
  line/.style={draw, -{Stealth[]}, thick}
]

  \node[box] (start) at (0,0) {};

  \node[box] (pref) at (3,2) {};
  \node[box] (recc) at (3,-2) {};

  \draw[line] (start) -- (pref);
  \draw[line] (start) -- (recc);

  \node at (2.65,0.75) {\tiny $\mathbb{P}_c(\omega_1=0)=1-\mu(c,k)$};
  \node at (2.35,-0.75) {\tiny $\mathbb{P}_c(\omega_1=1)=\mu(c,k)$};

  \node at (3,2.45) {\scriptsize 0};
  \node at (3,-1.45) {\scriptsize 1};

  \node[box] (j_pref) at (7.25,2) {};
  \node[box] (j_recc) at (7.25,-2) {};

  \draw[line] (pref) -- (j_pref);
  \draw[line] (recc) -- (j_recc);

  \node at (7.25,2.45) {\scriptsize $j$};
  \node at (7.25,-1.45) {\scriptsize $j$};

  \node at (5.1,2.3) {\tiny $\mathbb{P}_c(\omega_2=j\mid \omega_1=0)=L_{cj}(k)$};
  \node at (5.1,-1.7) {\tiny $\mathbb{P}_c(\omega_2=j\mid \omega_1=1)=x_{cj}(k)$};

  \node[box] (one_pref) at (12.5,2) {};
  \node[box] (zero_pref) at (7.25,0.5) {};

  \draw[line] (j_pref) -- (one_pref);
  \draw[line] (j_pref) -- (zero_pref);

  \node at (12.8,2) {\scriptsize 1};
  \node at (7.5,0.5) {\scriptsize 0};

  \node at (10,2.3) {\tiny $\mathbb{P}_c(\omega_3=1\mid \omega_1=0,\omega_2=j)=\gamma_p(k)$};
  \node at (9.5,1.25) {\tiny $\mathbb{P}_c(\omega_3=0\mid \omega_1=0,\omega_2=j)=1-\gamma_p(k)$};

  \node[box] (one_recc) at (12.5,-2) {};
  \node[box] (zero_recc) at (7.25,-3.5) {};

  \draw[line] (j_recc) -- (one_recc);
  \draw[line] (j_recc) -- (zero_recc);

  \node at (12.8,-2) {\scriptsize 1};
  \node at (7.5,-3.5) {\scriptsize 0};

  \node at (10,-1.65) {\tiny $\mathbb{P}_c(\omega_3=1\mid \omega_1=1,\omega_2=j)=\gamma_r(k)$};
  \node at (9.5,-2.75) {\tiny $\mathbb{P}_c(\omega_3=0\mid \omega_1=1,\omega_2=j)=1-\gamma_r(k)$};

  \draw[decorate, decoration={brace, amplitude=5pt}]
    (0,3.15) -- (3,3.15)
    node[midway, yshift=0.40cm] {\scriptsize Step 1: Recommendation};

  \draw[decorate, decoration={brace, amplitude=5pt}]
    (3.6,3.15) -- (7.4,3.15)
    node[midway, yshift=0.40cm] {\scriptsize Step 2: Repositioning};

  \draw[decorate, decoration={brace, amplitude=5pt}]
    (7.8,3.15) -- (12.9,3.15)
    node[midway, yshift=0.40cm] {\scriptsize Step 3: Allocation};

\end{tikzpicture}%
}
\caption{This figure illustrates the sequential decision-making process of a taxi driver $c \in \mathcal{C}$, showing the driver's decision to accept or reject a system recommendation, the subsequent movement to a region, and the allocation outcome.}
\label{fig:decision_tree}
\end{figure}

\subsection{Taxi Driver's Repositioning Decision-Making Model}

We consider a discrete-time setting with time steps indexed by \(k \in \mathcal{K}\), where each time step has duration \(H\). The service area is represented as a graph whose nodes correspond to service regions. Let \(\mathcal{R}\) denote the set of regions, and let \(\mathcal{C}\) denote the set of idle taxi drivers available for repositioning. Passengers arrive at graph nodes, where they are matched with available drivers and transported to other nodes corresponding to their destination regions. For each driver-region pair \((c,j) \in \mathcal{C}\times\mathcal{R}\), let \(\mathcal{T}_{cj}(k)\) denote the travel time required for driver \(c\) to reposition to region \(j\) in time step \(k\). Region \(j\) is said to be \emph{reachable} by driver \(c\) within one time-step if \(\mathcal{T}_{cj}(k) \leq H\). Idle drivers receive repositioning recommendations from the system in each time step $k$. A driver may either accept the recommendation and move to the recommended region or reject it and move according to their own repositioning preference. After arriving at the selected region, the driver is either matched with a passenger (allocated) or remains idle (unallocated).

This sequential decision-making process for driver \(c \in \mathcal{C}\) is modelled as a random experiment with a probability space \( (\Omega, \mathcal{F}, \mathbb{P}_c) \) and is shown in \cref{fig:decision_tree}. The sample space of the random experiment is \( \Omega = \{0, 1\} \times \mathcal{R} \times \{0, 1\} \), where an outcome is a tuple \( \omega = (\omega_1, \omega_2, \omega_3) \in \Omega \). Here, \( \omega_1 = 0 \) indicates that the driver rejected the system recommendation, while \( \omega_1 = 1 \) indicates that the driver accepted the system recommendation. Additionally, \( \omega_2 = j \) implies that the driver moved to region \( j \in \mathcal{R} \). Finally, \( \omega_3 = 1 \) implies that the driver was allocated a passenger, whereas \( \omega_3 = 0 \) indicates that the driver remained idle. The \(\sigma\)-algebra, of the sample space is denoted as \(\mathcal{F}\). For brevity, the following notation is used: \(\{\omega_1 = a\} = \{(\omega_1, \omega_2, \omega_3) \in \Omega \mid \omega_1 = a\}\). Similarly, this notation can be extended to other combinations of \(\omega_1\), \(\omega_2\), and \(\omega_3\) to represent events involving multiple conditions. The probability measure is defined such that \( \mathbb{P}_c(\{\omega_1 = 1\}) \) represents the probability of the driver accepting the system recommendation, while \( \mathbb{P}_c(\{\omega_1 = 0\}) \) represents the probability of the driver rejecting the system recommendation. Furthermore, \( \mathbb{P}_c(\{\omega_2 = j\} \mid \{\omega_1 = 0\}) \) denotes the probability of the driver moving to destination \( j \in \mathcal{R} \) given that the driver has rejected the system recommendation, and \( \mathbb{P}_c(\{\omega_2 = j\} \mid \{\omega_1 = 1\}) \) denotes the probability of the driver moving to destination \( j \in \mathcal{R} \) given that the driver has accepted the system recommendation. Finally, the probability of the driver getting allocated is defined such that if the driver rejected the system recommendation (i.e., repositioning according to own preference instead) and moved to region \( j \in \mathcal{R} \), the probability that the driver remains idle is \( \mathbb{P}_c(\{\omega_3 = 0\} \mid \{\omega_1 = 0\}, \{\omega_2 = j\}) = 1 - \gamma_p(k) \), and the probability that the driver gets allocated is \( \mathbb{P}_c(\{\omega_3 = 1\} \mid \{\omega_1 = 0\}, \{\omega_2 = j\}) = \gamma_p(k) \). Similarly, if the driver accepted the system recommendation and moved to region \( j \in \mathcal{R} \), the probability that the driver remains idle is \( \mathbb{P}_c(\{\omega_3 = 0\} \mid \{\omega_1 = 1\}, \{\omega_2 = j\}) = 1 - \gamma_r(k) \), and the probability that the driver gets allocated is \( \mathbb{P}_c(\{\omega_3 = 1\} \mid \{\omega_1 = 1\}, \{\omega_2 = j\}) = \gamma_r(k) \). 

\begin{table}[t]
\centering
\footnotesize
\renewcommand{\arraystretch}{1.3}
\setlength{\tabcolsep}{3pt}
\begin{tabular}{p{0.22\columnwidth} p{0.72\columnwidth}}
\hline
\textbf{Symbol} & \textbf{Description} \\
\hline

$\mathcal{C}, \mathcal{R}, \mathcal{K}$ & Sets of drivers, regions, and time steps \\
$c, j, k$ & Indices for driver $c$, region $j$, and time step $k$ \\
$H$ & Planning horizon, i.e., length of time step $k$ \\
$\omega=(\omega_1,\omega_2,\omega_3)$ & Driver outcome: acceptance, region choice, and allocation \\

$\mathcal{T}_{cj}(k)$ & Expected travel time from driver $c$'s location to region $j$ at time step $k$ \\

$\mu(c,k)$ & Probability that driver $c$ accepts the system recommendation at time step $k$ \\
$\hat{\mu}(c,k)$ & Platform's estimate of the driver's recommendation acceptance probability $\mu(c,k)$\\
$\alpha(c,k),\beta(c,k)$ & Beta confidence estimate parameters for driver $c$ at time step $k$ \\
$\epsilon_1(c),\epsilon_0(c)$ & Sensitivity parameters for confidence update\\
$y(c,k)$ & Binary allocation outcome of driver $c$ at time step $k$ \\
$\gamma_p(k),\gamma_r(k)$ & Allocation probabilities under preference and recommendation-driven moves \\

$r_j(k)$ & Feature vector for estimating preferences toward region $j$ at time step $k$ \\
$\sigma_c(\cdot)$ & Logistic function to estimate driver c's preference for region $j$ at time step $k$ \\
$L_{cj}(k)$ & Preference probability of driver $c$ for region $j$ at time step $k$ \\
$x_{cj}(k)$ & Recommendation for driver $c$ to reposition to region $j$ at time step $k$ (Binary)\\
$\mathbf{Y_{cj}(k)}$ & Bernoulli r.v. indicating whether driver $c$ arrives in region $j$ at time step $k$ \\
$P_{cj}(k)$ & Arrival probability of driver $c$ in region $j$ at time step $k$ \\

$\mathbf{S_j(k)}$ & Poisson-binomial r.v representing supply in region $j$ at time step $k$ \\
$\pi_j^{x(k)}(\cdot)$ & PMF of the realized supply $\mathbf{S_j(k)} \in \mathbb{Z}_+$ \\
$\mathbf{D_j(k)}$ & Poisson r.v representing demand in region $j$ at time step $k$ \\
$q_j^{k}(\cdot)$ & PMF of demand $\mathbf{D_j(k)} \in \mathbb{Z}_+$ \\
$\lambda_j(k)$ & Expected demand in region $j$ at time step $k$ \\
$z_j(k)$ & Expected served demand in region $j$ at time step $k$ \\

$\mathcal{X}(k)$ & Feasible recommendation set at time step $k$ \\

\hline
\end{tabular}
\end{table}

\subsection{Taxi Driver's Preference to Reposition to a Region}
\label{sec:driver-pref}

If driver \(c \in \mathcal{C}\) rejects the system recommendation, i.e., \(\omega_1=0\), the driver repositions according to their own regional preference. The probability that driver \(c\) chooses region \(j \in \mathcal{R}\) at time step \(k\) is
\begin{equation}
\mathbb{P}_c\bigl(\{\omega_2=j\}\mid \{\omega_1=0\}\bigr)
=
L_{cj}(k),
\end{equation}
where \(L_{cj}(k)\in[0,1]\) is given by
\begin{equation}
\label{eqn:preference}
L_{cj}(k)
=
\frac{\sigma_c(r_j(k))}
{\sum_{i\in\mathcal{R}}\sigma_c(r_i(k))},
\qquad
\sigma_c(r_j(k))
=
\frac{1}{1+\exp\{-\theta_c^\top r_j(k)\}} .
\end{equation}

Here, \(r_j(k)\) denotes the feature vector associated with region \(j\) at time step \(k\), and \(\sigma_c(\cdot)\) maps these features to an attractiveness score for driver \(c\). The parameter vector \(\theta_c\) is driver-specific and is learned from the historical repositioning decisions of driver \(c\).

If driver \(c\) accepts the system recommendation, i.e., \(\omega_1=1\), the driver moves to the recommended region. In this case, the destination choice is deterministic:
\begin{equation}
\mathbb{P}_c\bigl(\{\omega_2=j\}\mid \{\omega_1=1\}\bigr)
=
x_{cj}(k),
\end{equation}
where
\begin{equation}
\label{eqn:repositioning_decision_variable}
x_{cj}(k)
=
\begin{cases}
1, & \text{if driver \(c\) is issued a recommendation for region \(j\) at time step \(k\),}\\
0, & \text{otherwise.}
\end{cases}
\end{equation}

The variable \(x_{cj}(k)\in\{0,1\}\) is the system's repositioning decision: each driver receives exactly one recommendation, and any recommended region must be reachable within the planning horizon:
\begin{equation}
\label{eqn:allocation_reachability_constraint}
\sum_{j\in\mathcal{R}} x_{cj}(k) = 1,\ \forall c\in\mathcal{C},
\qquad
x_{cj}(k)\bigl(\mathcal{T}_{cj}(k)-H\bigr) \leq 0,\ \forall c\in\mathcal{C},\ j\in\mathcal{R}.
\end{equation}

\subsection{Taxi Drivers' Confidence in the Recommender System}
\label{sec:driver-conf}
The probability that taxi driver \(c \in \mathcal{C}\) accepts a system recommendation at time step \(k\) is denoted by \(\mu(c,k)\in(0,1)\). This acceptance probability evolves over time as the driver repeatedly interacts with the platform and observes recommendation outcomes.

Since the true acceptance probability \(\mu(c,k)\) is not directly observable by the platform, the recommender system instead maintains an estimate, denoted by \(\hat{\mu}(c,k)\). Hereafter, we refer to \(\hat{\mu}(c,k)\) as the driver's confidence in the recommender system. We characterize the evolution of this confidence estimate through a set of minimal structural properties that any plausible trajectory \(\{\hat{\mu}(c,k)\}\) should satisfy.

\subsubsection{Properties of the Driver Confidence Dynamics}

Let \(\{\hat{\mu}(c,k)\}_{k\ge0}\subset (0,1)\) denote the confidence trajectory of driver \(c\in\mathcal{C}\).
Each epoch generates a reward signal \(y(c,k)\in\{0,1\}\), where \(y(c,k)=1\) denotes a successful passenger allocation and \(y(c,k)=0\) denotes that the driver remains idle.
We impose the following minimal and behaviorally natural properties:

\begin{itemize}

    \item[(A1)] Evidence Direction Consistency:
    Confidence moves monotonically in the direction of the observed outcome:
    \begin{equation}
        \begin{aligned}
        y(c,k)=1 \Rightarrow \hat{\mu}(c,k+1)\ge \hat{\mu}(c,k),\\
        y(c,k)=0 \Rightarrow \hat{\mu}(c,k+1)\le \hat{\mu}(c,k).            
        \end{aligned}
    \end{equation}

    \item[(A2)] Convergence under Sustained Evidence:
    Under persistent identical outcomes, the confidence trajectory converges:
    \begin{equation}
    \begin{aligned}
        &y(c,k)=1 \ \forall k\ge K \Rightarrow \lim_{k\to\infty}\hat{\mu}(c,k)\ \text{exists}, \\
        &y(c,k)=0 \ \forall k\ge K \Rightarrow \lim_{k\to\infty}\hat{\mu}(c,k)\ \text{exists}.
    \end{aligned}
    \end{equation}

\end{itemize}

These assumptions define the admissible class of confidence trajectories.

\subsubsection{Taxi Drivers' Confidence Dynamics Model}
\begin{figure}[tbh]
    \centering
    \includegraphics[width=0.5\linewidth]{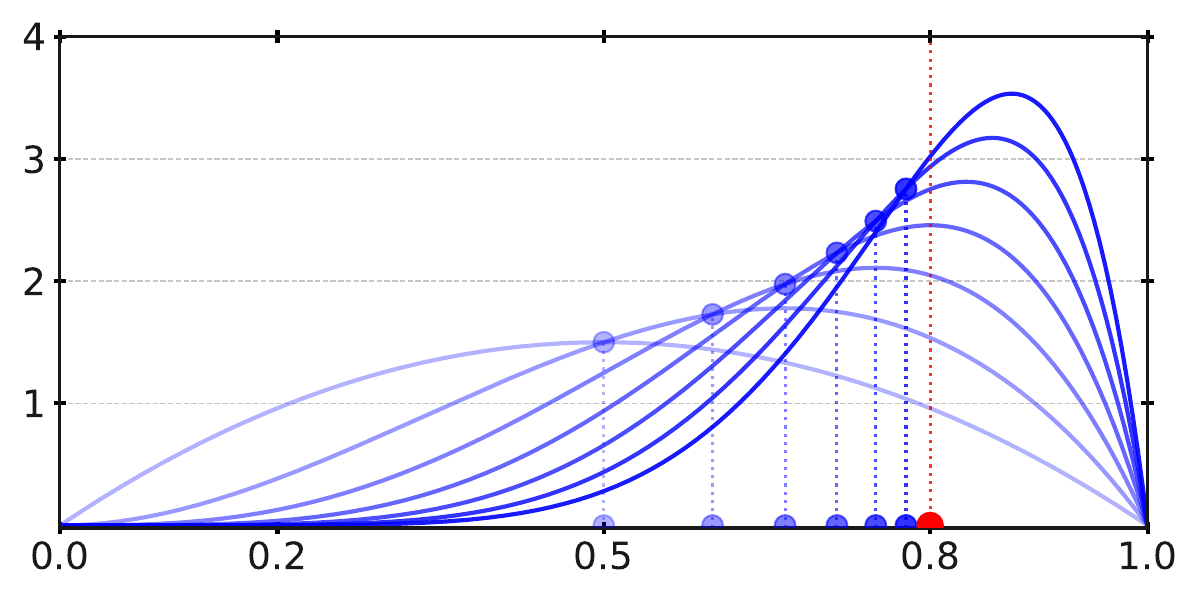}
    \caption{Evolution of the platform's estimate of the driver's acceptance probability under sustained evidence.
    Starting from a symmetric prior, repeated successes shift the mean confidence upward and reduce the variance of the Beta distribution.}
    \label{fig:beta distribution}
\end{figure}

We model the platform's estimate of the driver's acceptance probability using a \emph{Beta confidence model}, parameterized by the Beta distribution parameters \((\alpha(c,k),\beta(c,k))\). The corresponding confidence estimate is given by the mean of this distribution as shown in \cref{fig:beta distribution}:
\begin{equation}\label{eqn:confidence}
\hat{\mu}(c,k) 
= \mathbb{E}[\mathrm{Beta}(\alpha(c,k),\beta(c,k))]
= \frac{\alpha(c,k)}{\alpha(c,k)+\beta(c,k)}.
\end{equation}

\noindent Upon observing the outcome \(y(c,k)\in\{0,1\}\), the parameters are updated as
\begin{equation}\label{eq:alpha_beta_update}
\begin{aligned}
&\alpha(c,k+1) = \alpha(c,k) + \epsilon_1(c) \, y(c,k)\\
&\beta(c,k+1)  = \beta(c,k) + \epsilon_0(c) \, [1-y(c,k)],    
\end{aligned}
\end{equation}
where \(\epsilon_1(c),\epsilon_0(c) \in [0, \infty)\) quantify the sensitivity of the driver’s belief to successful and failed recommendations.
Larger \(\epsilon_1(c)\) produce stronger confidence increases after success, while larger \(\epsilon_0(c)\) produce stronger decreases after failure.

\subsubsection{Behavioral Consistency of the Beta confidence model}

The following result shows that the update rule~\eqref{eq:alpha_beta_update} always produces confidence trajectories satisfying the properties~(A1)–(A2).

\begin{theorem}[Behavioural Consistency]\label{thm:consistency}
For any reward sequence \(\{y(c,k)\}_{k\ge0}\subset\{0,1\}\) and any nonnegative update magnitudes \(\epsilon_1(c),\epsilon_0(c)\), 
the confidence trajectory \(\{\hat{\mu}(c,k)\}_{k\ge0}\) generated by the Beta confidence model satisfies properties \emph{(A1)} and \emph{(A2)}.
\end{theorem}

\begin{proof}
Fix a driver \(c\) and suppress the index for simplicity.  
Let \(\xi(k)=\alpha(k)+\beta(k)\) and \(\hat{\mu}(k)=\alpha(k)/\xi(k)\).

\emph{(A1)}  
If \(y(k)=1\) then
\[
\hat{\mu}(k+1)-\hat{\mu}(k)
= \frac{\epsilon_1\,\beta(k)}{\xi(k)\,\bigl(\xi(k)+\epsilon_1\bigr)} \ge 0,
\]
and if \(y(k)=0\) then
\[
\hat{\mu}(k+1)-\hat{\mu}(k)
= -\frac{\epsilon_0\,\alpha(k)}{\xi(k)\,\bigl(\xi(k)+\epsilon_0\bigr)} \le 0.
\]
Thus confidence moves monotonically in the direction of the outcome.

\emph{(A2)}
Suppose \(y(k)=1\) for all \(k\ge K\),
\[
\alpha(k)=\alpha(K)+(k-K)\epsilon_1,
\qquad
\beta(k)=\beta(K).
\]
Therefore,
\[
\hat{\mu}(k)
=
\frac{\alpha(K)+(k-K)\epsilon_1}
     {\alpha(K)+\beta(K)+(k-K)\epsilon_1}.
\]
If \(\epsilon_1\in(0,\infty)\), then \(\hat{\mu}(k)\to 1\) as \(k\to\infty\). 
If \(\epsilon_1=0\), then \(\hat{\mu}(k)=\hat{\mu}(k)\) for all \(k\ge K\). Hence, the limit exists.

Similarly, if \(y(k)=0\) for all \(k\ge K\), then
\[
\alpha(k)=\alpha(K),
\qquad
\beta(k)=\beta(K)+(k-K)\epsilon_0,
\]
and
\[
\hat{\mu}(k)
=
\frac{\alpha(K)}
     {\alpha(K)+\beta(K)+(k-K)\epsilon_0}.
\]
If \(\epsilon_0\in(0, \infty)\), then \(\hat{\mu}(k)\to 0\) as \(k\to\infty\). If \(\epsilon_0=0\), then \(\hat{\mu}(k)=\hat{\mu}(k)\) for all \(k\ge K\). Thus, the trajectory converges under sustained evidence.

Thus the Beta confidence model satisfies \emph{(A1)} and \emph{(A2)}.
\end{proof}

\begin{definition}[Arrival]
For a driver \(c\) standing idle, arrival is defined as the event that the driver repositions to a reachable region \(j\) within time step \(k\) either by adhering to the system recommendation or by following their own repositioning preference. The estimated arrival probability is defined as
\begin{equation}
\label{eqn:arrival_probability}
P_{cj}(k)
=
\underbrace{\hat{\mu}(c,k)\,x_{cj}(k)}_{\text{adhering to recommendation}}
+
\underbrace{\bigl(1-\hat{\mu}(c,k)\bigr)\,L_{cj}(k)}_{\text{following own preference}}.
\end{equation}
Here, \(P_{cj}(k)\) is the estimated probability that driver \(c\) arrives in region \(j\) at time step \(k\), determined jointly by the driver confidence \(\hat{\mu}(c,k)\), the preference probability \(L_{cj}(k)\), and the recommendation decision \(x_{cj}(k)\). For non-reachable regions \(l\), we define the estimated arrival probability as \(P_{cl}(k)=0\).
\end{definition}

Let \(\mathbf{Y_{cj}(k)}\) be the Bernoulli random variable indicating whether driver \(c \in \mathcal{C}\) \emph{arrives} in region \(j \in \mathcal{R}\) at time step \(k\). Then,
\begin{equation}\label{eqn: arrival-random-variable}
\mathbf{Y_{cj}(k)}
=
\begin{cases}
1, & \text{with probability } P_{cj}(k),\\
0, & \text{with probability } 1-P_{cj}(k).
\end{cases}
\end{equation}

\subsection{Adherence Aware Vehicle Rebalancing (AAVR) Model}
\label{sec:reb-recc}
The proposed model generates repositioning recommendations for idle taxi drivers with the objective of minimizing the expected mismatch between passenger demand and realized taxi supply over the planning horizon. It explicitly accounts for uncertainty in driver adherence, influenced by both the driver’s confidence in the recommender system and their individual preference for the suggested destination.  

\subsubsection{Realized Taxi Supply}
\label{sec:supp-dis}

Let \(\mathbf{S}_{\mathbf{j}}(\mathbf{k})\) denote the number of taxi drivers available in region \(j \in \mathcal{R}\) after repositioning at time step \(k \in \mathcal{K}\). It is obtained by aggregating the arrival decisions \(\mathbf{Y}_{\mathbf{c}\mathbf{j}}(\mathbf{k})\) of all drivers:
\begin{equation}
\label{eqn:supply}
\mathbf{S}_{\mathbf{j}}(\mathbf{k})
=
\sum_{c \in \mathcal{C}} \mathbf{Y}_{\mathbf{c}\mathbf{j}}(\mathbf{k}).
\end{equation}
Since the random variables \(\mathbf{Y_{cj}(k)}\) are independent Bernoulli random variables defined in \eqref{eqn: arrival-random-variable}, the total taxi count \(\mathbf{S_j(k)}\) follows a Poisson binomial distribution with probability mass function \(\pi_j^{x(k)}(\cdot)\), given by
\begin{equation}
\pi_j^{x(k)}(s):=\mathbb{P}\bigl(\mathbf{S_j(k)}=s\bigr)
=
\sum_{\substack{C_s \subseteq \mathcal{C} \\ |C_s|=s}}
\left(
\prod_{c \in C_s} P_{cj}(k)
\prod_{c \notin C_s} \bigl(1-P_{cj}(k)\bigr)
\right).
\end{equation}
where \(C_s \subseteq \mathcal{C}\) denotes a subset of drivers containing exactly \(s\) elements. The mean and variance of \(\mathbf{S_j(k)}\) are:
\begin{equation}
\mathbb{E}[\mathbf{S_j(k)}]
=
\sum_{c\in\mathcal{C}} P_{cj}(k),
\qquad
\mathrm{Var}(\mathbf{S_j(k)})
=
\sum_{c\in\mathcal{C}} P_{cj}(k)\bigl(1-P_{cj}(k)\bigr).
\end{equation}

\subsubsection{Expected Demand and Inter-Region Travel Time}
\label{sec:demand-travel-time}
We model regional demand during each planning horizon of length \(H\) as a Poisson random variable \(\mathbf{D_j(k)} \in \mathbb{Z}_{\ge 0}\). Let \(q_j^k(\cdot)\) denote its probability mass function, given by
\begin{equation}
q_j^k(d)
:=
\mathbb{P}\!\left(\mathbf{D_j(k)}=d\right)
=
\frac{e^{-\lambda_j(k)}\lambda_j(k)^d}{d!},
\qquad
\lambda_j(k)=g_j\bigl(\tilde{D}_j(k)\bigr).
\label{eqn:expected_demand}
\end{equation}
where \(\lambda_j(k)\) denotes the expected number of passenger requests in region \(j\) during time step \(k\), and \(g_j(\cdot)\) is a regression model that predicts \(\lambda_j(k)\) from the feature vector \(\tilde{D}_j(k)\). The feature vector \(\tilde{D}_j(k)\) comprises lagged demand observations and temporal features, including the hour of the day, day of the week, and day of the month.
Similarly, for each origin-destination pair \((i,j)\in\mathcal{R}\times\mathcal{R}\), we model the inter-region travel time as
\begin{equation}
\mathbf{T_{ij}}(k)\sim\mathcal{N}_{+}\bigl(\tau_{ij}(k),\epsilon_{ij}^2(k)\bigr),
\qquad
\tau_{ij}(k)=h\bigl(\tilde{T}_{ij}(k)\bigr),
\label{eqn:expected_travel_time}
\end{equation}
where \(\mathcal{N}_{+}\) denotes the normal distribution truncated to the nonnegative real line, \(\tau_{ij}(k)\) is the expected travel time from region \(i\) to region \(j\) during time step \(k\), and \(h(\cdot)\) is a regression model that predicts \(\tau_{ij}(k)\) from the feature vector \(\tilde{T}_{ij}(k)\). The feature vector \(\tilde{T}_{ij}(k)\) comprises historical travel times from region \(i\) to region \(j\), the distance between the two regions, the hour of the day, and the day of the week. The variance parameter \(\epsilon_{ij}^2(k)\) is estimated from the residuals of the regression model on a held-out validation dataset, where the residuals are computed as the differences between the observed and predicted travel times.

For a driver \(c\in\mathcal{C}\) located in region \(i\in\mathcal{R}\) at time step \(k\), we define the expected driver-region repositioning time to region \(j\in\mathcal{R}\) as
\begin{equation}
\label{eqn:travel_time_for_driver}
\mathcal{T}_{cj}(k):=\tau_{ij}(k).
\end{equation}

\subsubsection{Adherence-Aware Vehicle Rebalancing (AAVR) Problem}

For each time step \(k \in \mathcal{K}\), the admissible repositioning decisions are
\begin{equation}
\label{eqn:feasible_set}
\begin{aligned}
\mathcal{X}(k)
=
\Bigl\{
x(k)\in\{0,1\}^{|\mathcal{C}|\times|\mathcal{R}|}
\;\Big|\;
&\sum_{j\in\mathcal{R}} x_{cj}(k)=1,\ \forall c\in\mathcal{C},\\
&x_{cj}(k)\bigl(\mathcal{T}_{cj}(k)-H\bigr)\le 0,\ 
\forall c\in\mathcal{C},\ j\in\mathcal{R}
\Bigr\}.
\end{aligned}
\end{equation}
The first constraint assigns exactly one recommended region to each driver, while the second ensures that a recommended region is reachable within the planning horizon.

For a given decision \(x(k)\), the realized supply and demand in region \(j\in\mathcal{R}\) satisfy
\begin{equation}
\label{eqn:supply_demand_distribution}
\mathbf{S_j(k)}
=
\sum_{c\in\mathcal{C}} \mathbf{Y_{cj}(k)},
\qquad
\mathbf{Y_{cj}(k)}\sim \mathrm{Bernoulli}\bigl(P_{cj}(k)\bigr),
\qquad
\mathbf{D_j(k)}
\sim
\mathrm{Poisson}\bigl(\lambda_j(k)\bigr),
\end{equation}
where \(P_{cj}(k)\) is defined in \eqref{eqn:arrival_probability}. Assuming \(\mathbf{S_j(k)}\) and \(\mathbf{D_j(k)}\) are independent, we define the \emph{cost-of-deviation} at time step \(k\) as the expected absolute mismatch between demand and realized supply:
\begin{definition}[Cost of deviation]
\label{def:cost_of_deviation}
The \emph{cost-of-deviation} measures the expected mismatch between passenger demand and realized taxi supply across all regions. The objective penalizes both under-supply and over-supply. While under-supply increases passenger waiting time, over-supply may lead to idle and unallocated drivers, which can reduce driver confidence in the recommender system over time.
\begin{equation}
\label{eqn:cost_of_deviation}
\mathbb{E}\!\left[
\sum_{j\in\mathcal{R}}
\left|
\mathbf{D_j(k)}-\mathbf{S_j(k)}
\right|
\right]
=
\sum_{j\in\mathcal{R}}
\sum_{d=0}^{\infty}
\sum_{s=0}^{|\mathcal{C}|}
|d-s|\,
\pi_j^{x(k)}(s)\,
q_j^{k}(d).
\end{equation}
\end{definition}

The AAVR problem is then given by the following optimization problem, where the decision \(x(k)\) determines the recommendation provided to each driver, the probabilities \(P_{cj}(k)\) determine the induced arrival probabilities, and the objective minimizes the expected \emph{cost-of-deviation} across all regions.

\begin{problem}[Adherence-Aware Vehicle Rebalancing (AAVR) Problem]
\label{prob:AAVR}
At each time step \(k\in\mathcal{K}\), given the expected demand \(\lambda_j(k)\), driver confidence \(\hat{\mu}(c,k)\), preference probability \(L_{cj}(k)\), and repositioning time \(\mathcal{T}_{cj}(k)\), determine a recommendation decision \(x(k)\in\mathcal{X}(k)\) that solves
\begin{equation}
\label{eqn:aavr_problem}
\begin{aligned}
\min_{x(k)\in\mathcal{X}(k)} \quad
& \sum_{j\in\mathcal{R}}
\sum_{d=0}^{\infty}
\sum_{s=0}^{|\mathcal{C}|}
|d-s|\,
\pi_j^{x(k)}(s)\,
q_j^{k}(d)
\\
\text{s.t.}\quad
& \mathcal{X}(k)
=
\Bigl\{
x(k)\in\{0,1\}^{|\mathcal{C}|\times|\mathcal{R}|}
\ \Big|\ 
\sum_{j\in\mathcal{R}}x_{cj}(k)=1,\ 
x_{cj}(k)(\mathcal{T}_{cj}(k)-H)\le 0
\Bigr\},\\
\end{aligned}
\end{equation}
\end{problem}

Evaluating the distribution of \(\pi_j^{x(k)}(\cdot)\) requires summation over all subsets of drivers, resulting in \(2^{|\mathcal{C}|}\) terms. Moreover, since the probabilities \(P_{cj}(k)\) depend on the binary decision variables \(x_{cj}(k)\), the optimization problem is nonlinear and combinatorial. Directly solving this problem is computationally intractable. 

This structure motivates the next section, where we derive theoretical bounds on the \emph{cost-of-deviation} objective (\ref{eqn:cost_of_deviation}) and use them to construct a tractable model, suitable for real-time vehicle rebalancing applications, where recommendations must be computed and disseminated within each planning horizon.

\section{Theoretical Bounds for AAVR Objective Function}
\label{sec:theoretical}
This section derives theoretical bounds for the \emph{Adherence-Aware Vehicle Rebalancing (AAVR)} problem introduced in \cref{prob:AAVR}. 
We first reduce the analysis to a single target region and bound the \emph{cost-of-deviation} objective using the expected supply and the variance of the supply. 
These bounds separate the problem into two components: selecting a target expected supply and controlling the variance of realized supply around that target. 
We then derive the target expected supply through a median-matching argument and characterize the variance-minimizing recommendation structure at fixed expected supply. 
The resulting decomposition provides the basis for the tractable AAVR model.

\subsection{Preliminaries}
\begin{definition}[Convex order]
\label{def:convex-order}
Let $X,Y$ be real-valued random variables with $\mathbb{E}[X]=\mathbb{E}[Y]$. We write $X \le_{cx} Y$ if
\begin{equation}
\mathbb{E}[\varphi(X)] \le \mathbb{E}[\varphi(Y)] \quad \text{for every convex function } \varphi.
\end{equation}
\end{definition}

\begin{definition}[Majorization]
\label{def:majorization}
For $p,q \in [0,1]^n$ with components sorted as $p_1 \ge \cdots \ge p_n$ and $q_1 \ge \cdots \ge q_n$, we write $p \succeq q$ ($p$ majorizes $q$) if
\begin{equation}
\sum_{c=1}^k p_c \ge \sum_{c=1}^k q_c \quad \text{for } k=1,\dots,n-1, \quad \sum_{c=1}^n p_c = \sum_{c=1}^n q_c.
\end{equation}
\end{definition}

\begin{definition}[Majorization-maximal element]
\label{def:maj-max}
A vector $p^\star \in [0,1]^n$ is \emph{majorization-maximal} if $\nexists\, q \in [0,1]^n$ such that $q \succeq p^\star$, i.e., $p^\star$ is not majorized by any other feasible vector (it need not majorize all others).
\end{definition}

\begin{lemma}[Majorization implies convex-order dominance]
\label{lem:maj-to-cx}
Let $p,q \in [0,1]^n$ satisfy $\sum_{c=1}^n p_c = \sum_{c=1}^n q_c$. 
If $p \succeq q$, then the Poisson--binomial sums
$S_p := \sum_{c=1}^n \mathrm{Bernoulli}(p_c)$ and
$S_q := \sum_{c=1}^n \mathrm{Bernoulli}(q_c)$ satisfy
$S_p \le_{cx} S_q$.
This result follows from \cite[Theorem~3.A.37]{shaked2007stochastic}.
\end{lemma}

\subsection{Single-Region Cost-of-Deviation Bounds}
\label{sec:single-region-bounds}
Consider a \emph{target region} \(j \in \mathcal{R}\), and let the set of drivers that can reach this target region within the planning horizon \(H\) be indexed by \(c \in \{1,\dots,n\}\). We also consider one particular time-step \(k \in \mathcal{K}\), so we drop \(j\) and \(k\) for notational convenience. 
Let \(\hat{\mu}(c)\in[0,1]\) denote the confidence of driver \(c\), and define \(\ell_c := (1-\hat{\mu}(c))L_c \in [0,1]\) and \(w_c := \hat{\mu}(c)\).
A recommendation plan \(x \in \{ 0, 1\}^{n}\) is a vector of binary recommendations \(x_c \in \{0,1\}\) to each driver, that yields the arrival probability,
\begin{equation}\label{eqn:arrival_prob_ss}
p_c(x_c) = \ell_c + w_c x_c.
\end{equation}
Let \(\mathbf{Y_c} \sim \mathrm{Bernoulli}\!\bigl(p_c(x_c)\bigr)\) be independent Bernoulli random variables indicating whether driver \(c\) arrives at the target region. The realized supply and its expectation are given by
\begin{equation}
\mathbf{S} := \sum_{c=1}^n \mathbf{Y_c}, 
\qquad 
m(x) := \mathbb{E}[\mathbf{S}] = \sum_{c=1}^n p_c(x_c),
\end{equation}
where \(\mathbf{S} \in \mathbb{Z}_+\) follows a Poisson--binomial distribution. We define the attainable interval of expected supply as follows,
\begin{equation}
L_0:=\sum_{c=1}^{n} \ell_c,\quad U_0:=\sum_{c=1}^{n} (\ell_c+w_c)=L_0+\sum_{c=1}^{n} w_c,    
\end{equation}
Let \(\mathbf{D} \in \mathbb{Z}_+\) denote a Poisson random variable representing demand in the target region during the planning horizon. We assume that the demand \(\mathbf{D}\)  is independent of supply \(\mathbf{S}\). Next, we analyze the structure of the deviation cost.
\begin{definition}[Cost of deviation (single region)]\label{def:cost_of_deviation_single}
Let \(\pi^{x}(s):=\mathbb{P}\big(\mathbf{S}=s\big)\) denote the Poisson-binomial PMF of the realized supply \(\mathbf{S}\). Let \(q(d):=\mathbb{P}(\mathbf{D}=d)\) denote the PMF of demand \(\mathbf{D}\), where \(\mathbf{D} \sim \mathrm{Poisson}(\lambda)\). The \emph{cost of deviation} is defined as
\begin{equation}\label{eq:cost_of_deviation_single}
\mathcal{J}(x) 
:= \mathbb{E}\!\left[|\mathbf{D} - \mathbf{S}|\right]
= \sum_{d=0}^{\infty} \sum_{s=0}^{n} |d - s|\; \pi^{x}(s)\, q(d).
\end{equation}
\end{definition}
To overcome the difficulty of directly minimizing \(\mathcal{J}(x)\), we next derive the bounds that expose the structure of the objective. These bounds show that the \emph{cost-of-deviation} is governed by the expected value, and the variance of the realized supply \(\mathbf{S}\). The first result establishes a lower bound that depends only on the expected supply, while the second result quantifies how far the true objective can deviate from this bound as a function of the variance of the realized supply.

\begin{lemma}[Expectation-based lower bound]
For any decision vector $x$,
\begin{equation}
\mathcal{J}(x) \ge \mathbb{E}_{\mathbf{D}}\!\left[|\mathbf{D} - m(x)|\right].
\end{equation}
\end{lemma}
\begin{proof}
By the law of iterated expectations,
\begin{equation}
\mathcal{J}(x) = \mathbb{E}_{\mathbf{D}}\!\left[\mathbb{E}_{\mathbf{S}}\!\left[|\mathbf{D} - \mathbf{S}| \mid \mathbf{D}\right]\right].
\end{equation}
For any $\mathbf{D}=d$, the map $s \mapsto |d-s|$ is convex, hence by Jensen's inequality,
\begin{equation}
\mathbb{E}_{\mathbf{S}}\!\left[|\mathbf{D}-\mathbf{S}| \mid \mathbf{D}=d\right] \ge \left|d - \mathbb{E}_{\mathbf{S}}\!\left[\mathbf{S} \mid \mathbf{D}=d\right]\right|.
\end{equation}
Using independence, $\mathbb{E}_{\mathbf{S}}\!\left[\mathbf{S} \mid \mathbf{D}=d\right]=m(x)$, so
\begin{equation}
\mathbb{E}_{\mathbf{S}}\!\left[|d-\mathbf{S}| \mid \mathbf{D}=d\right] \ge |d-m(x)|.
\end{equation}
Taking expectation over $\mathbf{D}$ yields the result.
\end{proof}

\begin{lemma}[Variance-based upper bound]
For any decision vector $x$,
\begin{equation}\label{eqn:upper-bound}
\mathcal{J}(x)
\le
\mathbb{E}_{\mathbf{D}}\!\left[|\mathbf{D} - m(x)|\right]
+
\sqrt{\mathrm{Var}(\mathbf{S})}.
\end{equation}
\end{lemma}

\begin{proof}
Using the triangle inequality,
\begin{equation}
|\mathbf{D}-\mathbf{S}|
=
|\mathbf{D}-m(x) + m(x)-\mathbf{S}|
\le
|\mathbf{D}-m(x)| + |\mathbf{S}-m(x)|.
\end{equation}
Taking expectation with respect to the joint randomness of $\mathbf{D}$ and $\mathbf{S}$, we obtain
\begin{equation}
\mathcal{J}(x)
=
\mathbb{E}_{\mathbf{D},\mathbf{S}}\!\left[|\mathbf{D}-\mathbf{S}|\right]
\le
\mathbb{E}_{\mathbf{D}}\!\left[|\mathbf{D}-m(x)|\right]
+
\mathbb{E}_{\mathbf{S}}\!\left[|\mathbf{S}-m(x)|\right].
\end{equation}
By Cauchy--Schwarz,
\begin{equation}
\mathbb{E}_{\mathbf{S}}\!\left[|\mathbf{S}-m(x)|\right]
\le
\sqrt{
\mathbb{E}_{\mathbf{S}}\!\left[(\mathbf{S}-m(x))^2\right]
}
=
\sqrt{\mathrm{Var}(\mathbf{S})},
\end{equation}
which proves the result.
\end{proof}

\noindent Combining the lower and upper bounds, we obtain
\begin{equation}\label{eqn:upper bound}
\underbrace{\mathbb{E}_{\mathbf{D}}\!\left[|\mathbf{D} - m(x)|\right]}_{\text{mean mismatch}}
\;\le\;
\mathcal{J}(x)
\;\le\;
\underbrace{\mathbb{E}_{\mathbf{D}}\!\left[|\mathbf{D} - m(x)|\right]}_{\text{mean mismatch}}
+
\underbrace{\sqrt{\mathrm{Var}(\mathbf{S})}}_{\text{dispersion}}.
\end{equation}

The upper bound defined above gives a moment-based surrogate for the \emph{cost-of-deviation}. While the original objective \(\mathbb{E}[|\mathbf{D}-\mathbf{S}|]\) requires evaluating the full Poisson--binomial distribution of \(\mathbf{S}\), the bound avoids this distribution-level computation. This representation separates the \emph{cost-of-deviation} into a mean-mismatch term and a dispersion term. The former depends on the expected supply \(m(x)=\mathbb{E}[\mathbf{S}]\), while the latter depends on the supply variance \(\mathrm{Var}(\mathbf{S})\). 

We adopt a sequential approach for minimizing the upper bound to the \emph{cost-of-deviation} objective. We first select a target mean \(m^\star\) that minimizes the mismatch term \(\mathbb{E}_\mathbf{D}[|\mathbf{D} - m(x)|]\). Then, among all feasible decisions achieving this mean, we select the one that minimizes \(\mathrm{Var}(\mathbf{S})\). 

\subsection{Stage 1: Minimizing Mean-Mismatch: Target Expected Supply}
\label{sec:target-expected-supply}

We first determine the target expected supply by minimizing the mean-mismatch term in the \emph{cost-of-deviation} upper bound. The following result shows that the optimal target expected supply is obtained by matching the median demand, projected onto the attainable interval \([L_0,U_0]\).

\begin{theorem}[Median matching]
\label{thm:median matching}
The minimizer of the mean-mismatch term over the attainable expected-supply interval \([L_0,U_0]\) is
\begin{equation}
m^\star
= \min\!\bigl\{\max\{\operatorname{Median}(\mathbf{D}),\,L_0\},\,U_0\bigr\}.
\end{equation}
\end{theorem}

\begin{proof}
Let \(g(m) := \mathbb{E}[|\mathbf{D}-m|]\). The first stage solves \(\min_{m \in [L_0,U_0]} g(m)\). The function \(g\) is convex with subdifferential
\begin{equation}
\partial g(m) = [\,2F_{\mathbf{D}}(m^-)-1,\;2F_{\mathbf{D}}(m)-1\,].
\end{equation}
Thus \(0 \in \partial g(m)\) if and only if \(m\) is a median of \(\mathbf{D}\). Therefore, the constrained minimizer over \([L_0,U_0]\) is the projection of a median of \(\mathbf{D}\) onto this interval:
\begin{equation}\label{eqn:target}
m^\star
= \min\!\bigl\{\max\{\operatorname{Median}(\mathbf{D}),\,L_0\},\,U_0\bigr\}.
\end{equation}
\end{proof}

After selecting the target expected supply \(m^\star\), the remaining question is how to minimize \(\mathrm{Var}(\mathbf{S})\) at fixed mean \(\mathbb{E}[\mathbf{S}] = m^\star\).

\begin{remark}
In principle, at fixed mean, we can directly minimize the cost-of-deviation objective \(\mathcal{J}(x)\) instead of the variance term in the upper bound. It can be shown that \(\mathcal{J}(x)\) is minimized by any plan whose induced probability distribution is smaller than all other feasible distributions in convex order. For Poisson-binomial supplies, this can be obtained if the corresponding probability vector majorizes all other feasible probability vectors. However, such a globally majorizing vector need not exist under heterogeneous box constraints, i.e., driver-specific lower and upper probability bounds \(\ell_c \le p_c(x) \le u_c\). We provide the convex-order based sufficient condition and an example of non-existence in \ref{app:convex-order}. 
\end{remark}

Variance minimization at fixed mean and box constraints does not have a closed-form solution; however, we can provide the structure of the optimal solution. Additionally, we can show that the optimal leads to a majorization maximal arrival probability vector, i.e., no other feasible probability vector majorizes the variance maximization optimal.       

\subsection{Stage 2: Variance Minimization at Target Expected Supply}
\label{sec:fixed-expected-supply-variance}

Having selected the target expected supply \(m^\star\), we next choose among recommendation plans that achieve this target. Specifically, we consider the fixed-mean feasible set
\begin{equation}
\mathcal{X}_{m^\star}
=
\{x:\mathbb{E}[\mathbf{S}]=m^\star\}.
\end{equation}
For all \(x\in\mathcal{X}_{m^\star}\), the mean-mismatch term is fixed. Therefore, using the upper bound (\ref{eqn:upper-bound}), the remaining criterion is to control the variance of the realized supply. We thus select a recommendation plan by solving:
\begin{equation}
\min_{x\in\mathcal{X}_{m^\star}}
\mathrm{Var}(\mathbf{S}).
\end{equation}
We study a relaxed problem where we initially allow the recommendation to be \(0 \leq x_c \leq 1\), the result shows that the optimal can have at-most one \(x_c\) that is fractional. This result derives the solution structure, and a majorization maximal property using the Karush-Kuhn-Tucker (KKT) conditions.   

\begin{theorem}[Majorization-maximality and structure]
\label{thm:majorization-maximal}
Within the fixed-mean, box-constrained Bernoulli family
\begin{equation}
\mathbf{S}=\sum_{c=1}^{n}\mathrm{Bernoulli}\!\bigl(p_c(x)\bigr),\quad
0 \le \ell_c \le p_c(x) \le u_c \le 1,\quad
\sum_{c=1}^{n} p_c(x) = m^\star,
\end{equation}
let $x^\star$ minimize $\mathrm{Var}(\mathbf{S})$ and define $p^\star := p(x^\star)$. Then:

(i) $p^\star$ has the endpoint/one-fraction structure, i.e., at most one coordinate lies in $(\ell_c,u_c)$;

(ii) $p^\star$ is majorization-maximal on $\mathcal{P}_{m^\star} = \{q:\ 0 \le \ell_c \le q_c \le u_c \le 1,\ \sum_c q_c = m^\star\}$, i.e., $\nexists\, q \in \mathcal{P}_{m^\star}$ such that $q \succeq p^\star$.
\end{theorem}

\begin{proof}
Since $\sum_c p_c(x)=m^\star$ is fixed, we have
\begin{equation}
\mathrm{Var}(\mathbf{S}) = \sum_c p_c(x)(1-p_c(x)) = m^\star - \sum_c p_c(x)^2,
\end{equation}
hence minimizing $\mathrm{Var}(\mathbf{S})$ is equivalent to
\begin{equation}
\max_{p}\ \sum_c p_c^2 \quad \text{s.t.}\quad \sum_c p_c = m^\star,\ 0 \le \ell_c \le p_c \le u_c \le 1.
\end{equation}
Form the Lagrangian
\begin{equation}
\mathcal{L}(p,\lambda,\alpha,\beta) = \sum_c p_c^2 + \lambda\!\left(m^\star - \sum_c p_c\right) + \sum_c \alpha_c(p_c-\ell_c) + \sum_c \beta_c(u_c-p_c),
\end{equation}
with multipliers $\alpha_c,\beta_c \ge 0$. The KKT conditions are
\begin{equation}
2p_c - \lambda + \alpha_c - \beta_c = 0,\quad
\alpha_c(p_c-\ell_c)=0,\quad \beta_c(u_c-p_c)=0.
\end{equation}
Thus, for each coordinate $c$:
\begin{equation}
p_c \in (\ell_c,u_c) \Rightarrow \alpha_c=\beta_c=0 \Rightarrow p_c=\tfrac{\lambda}{2},
\end{equation}
\begin{equation}
p_c=\ell_c \Rightarrow \beta_c=0,\ \alpha_c \ge 0 \Rightarrow 2\ell_c \le \lambda,\qquad
p_c=u_c \Rightarrow \alpha_c=0,\ \beta_c \ge 0 \Rightarrow 2u_c \ge \lambda.
\end{equation}

Hence all interior coordinates must take the same value $\lambda/2 \in [0,1]$. If two indices $i,j$ were interior, then $p_i=p_j=\lambda/2$; consider the feasible perturbation $p_i=\tfrac{\lambda}{2}+\varepsilon,\ p_j=\tfrac{\lambda}{2}-\varepsilon$, preserving the sum and feasibility for sufficiently small $\varepsilon$. The objective changes by $2\varepsilon^2>0$, contradicting optimality. Therefore, at most one coordinate can be interior, establishing (i).

For (ii), note that $\phi(p)=\sum_c p_c^2$ is symmetric and strictly convex on $[0,1]^n$, hence strictly Schur-convex \cite[Def.~A.1]{marshall1979inequalities}. Thus,
\begin{equation}
q \succ p,\ q \neq p \ \Rightarrow\ \phi(q)>\phi(p).
\end{equation}
If there existed $q \in \mathcal{P}_{m^\star}$ such that $q \succ p^\star$, then
\begin{equation}
\sum_c q_c^2 > \sum_c (p^\star_c)^2,
\end{equation}
contradicting optimality. Hence $p^\star$ is majorization-maximal.
\end{proof}

\begin{remark}
The preceding analysis establishes the structural properties of the optimal recommendation policy for a single target region. We use these results to motivate the network-level optimization model by applying the median-matching and variance-minimization principles across all regions. The resulting formulation is therefore a tractable approximation of \cref{prob:AAVR}, obtained through a weighted multi-objective optimization based on the upper bound in \eqref{eqn:upper bound}.
\end{remark}

\subsection{Tractable Reformulation of AAVR Optimization Model}
Based on the structural insights established above, we formulate the following objectives to approximate the original AAVR problem.
\begin{enumerate}
\item \textit{Mean matching.}  
To align supply with demand, we follow the approach outlined in \cref{thm:median matching}. For Poisson demand distributions, the median is known to be close to the mean, especially as the rate increases. Motivated by this, we adopt a mean-matching approximation and introduce the expected-allocation variable
\begin{equation}
z_j(k) = \min\bigl(\mathbb{E}[\mathbf{D_j(k)}],\; \mathbb{E}[\mathbf{S_j(k)}]\bigr),
\end{equation}
and maximize \(\sum_{j} z_j(k) - \beta_0 \sum_{c,j} \mathcal{T}_{cj}(k)\, x_{cj}(k)\). This encourages matching expected supply to expected demand across regions. 

The travel-time penalty serves a dual purpose: it prioritizes nearby repositioning recommendations and penalizes excessive concentration of drivers in any single region, thereby mitigating oversupply. Together with the expected-allocation objective, it promotes mean matching across the network.

\item \textit{Variance minimization.}  
The dispersion term in the upper bound depends on the variance of the realized supply. For each region \(j\),
\begin{equation}
\mathrm{Var}(\mathbf{S_j(k)}) = \sum_{c \in \mathcal{C}} P_{cj}(k)\bigl(1 - P_{cj}(k)\bigr),
\end{equation}
where the arrival probability is
\[
P_{cj}(k) = \hat{\mu}(c,k)\,x_{cj}(k) + \bigl(1 - \hat{\mu}(c,k)\bigr)L_{cj}(k).
\]
Using the binary property \(x_{cj}^2(k) = x_{cj}(k)\), this expression can be rewritten as
\begin{equation}
P_{cj}(k)\bigl(1 - P_{cj}(k)\bigr)
=
\hat{\mu}(c,k)(1-\hat{\mu}(c,k))(1-2L_{cj}(k))\,x_{cj}(k)
+ \text{constant},
\end{equation}
which separates into a constant term and a linear term in \(x_{cj}(k)\). Consequently, the variance penalty can be incorporated directly into the objective as a linear function of the decision variables.
\end{enumerate}

Combining these two objectives yields the following tractable approximation of \cref{prob:AAVR}. The resulting optimization model is suitable for real-time vehicle rebalancing applications.

\begin{problem}[Tractable Reformulation of AAVR]
\label{prob:reb_opt}
At each time step \(k \in \mathcal{K}\), given expected demand \(\lambda_j(k)\), driver confidence \(\hat{\mu}(c,k)\), driver preference \(L_{cj}(k)\), and travel time \(\mathcal{T}_{cj}(k)\), determine recommendation decision \(x(k)\) and expected allocation \(z(k)\) that solves
\begin{equation}
\label{eqn:reb_opt_model}
\begin{aligned}
\max_{x(k), z(k)} \quad 
& \sum_{j \in \mathcal{R}} z_j(k)
- \beta_0 \sum_{c \in \mathcal{C} \, j \in \mathcal{R}} x_{cj}(k)\,\mathcal{T}_{cj}(k)
- \beta_1 \sum_{c \in \mathcal{C} \, j \in \mathcal{R}} 
\hat{\mu}(c,k)(1-\hat{\mu}(c,k))(1-2L_{cj}(k))\,x_{cj}(k) \\
\text{s.t.} \quad
& \sum_{j \in \mathcal{R}} x_{cj}(k) = 1 
\qquad \forall c \in \mathcal{C}, \\
& x_{cj}(k)\,(\mathcal{T}_{cj}(k) - H) \le 0 
\qquad \forall c \in \mathcal{C},\; \forall j \in \mathcal{R}, \\
& z_j(k) \le \lambda_j(k) 
\qquad \forall j \in \mathcal{R}, \\
& z_j(k) \le \sum_{c \in \mathcal{C}} 
\Big( \hat{\mu}(c,k)\,x_{cj}(k) + (1-\hat{\mu}(c,k))\,L_{cj}(k) \Big)
\qquad \forall j \in \mathcal{R}, \\
& x_{cj}(k) \in \{0,1\}, \quad z_j(k) \ge 0 
\qquad \forall c \in \mathcal{C},\; \forall j \in \mathcal{R}.
\end{aligned}
\end{equation}
\end{problem}

\subsubsection{Confidence Parameter Update:}
The confidence parameters for each driver \(c\) are updated based on their observed response to recommendations and allocation outcome:
\begin{enumerate}
\item If a driver accepts the recommendation (\(\omega_1=1\)), and \(y(c,k)\) is the allocation outcome, then the parameters are updated as:
\begin{equation}
\alpha_c(k+1) = \alpha_c(k) + \epsilon_1(c)\,y(c,k), 
\qquad 
\beta_c(k+1) = \beta_c(k) + \epsilon_0(c)\,\bigl(1 - y(c,k)\bigr).
\end{equation}

\item If the driver rejects the recommendation (\(\omega_1=0\)), no update is performed:
\begin{equation}
\alpha_c(k+1) = \alpha_c(k), 
\qquad 
\beta_c(k+1) = \beta_c(k).
\end{equation}
\end{enumerate}

\section{Experiments and Performance Evaluation}
\label{sec:results}
In this section, we discuss the experiments conducted to test the proposed adherence-aware vehicle rebalancing model in a large-fleet setting.
We explain the operational details of the proposed driver preference and confidence-aware taxi repositioning recommendation model and discuss the components that will become the key ingredients of a realistic test-bed used for simulation case studies. 
We develop a simulated taxi network based on \cite{donovan2014new} from New York City for the period of January to July 2010.\footnote{Specifically, we have used six months (Jan-June) of data from~\cite{donovan2014new} for training and one month (July) for testing the prediction models.}
To generate the road network, we leveraged an approach similar to our prior work~\citep{brar2021dynamic}, in which Manhattan city was partitioned into regions (taxi zones as defined by \cite{donovan2014new}). 
\begin{figure}[tbh]
    \centering
    \includegraphics[width=0.5\textwidth]{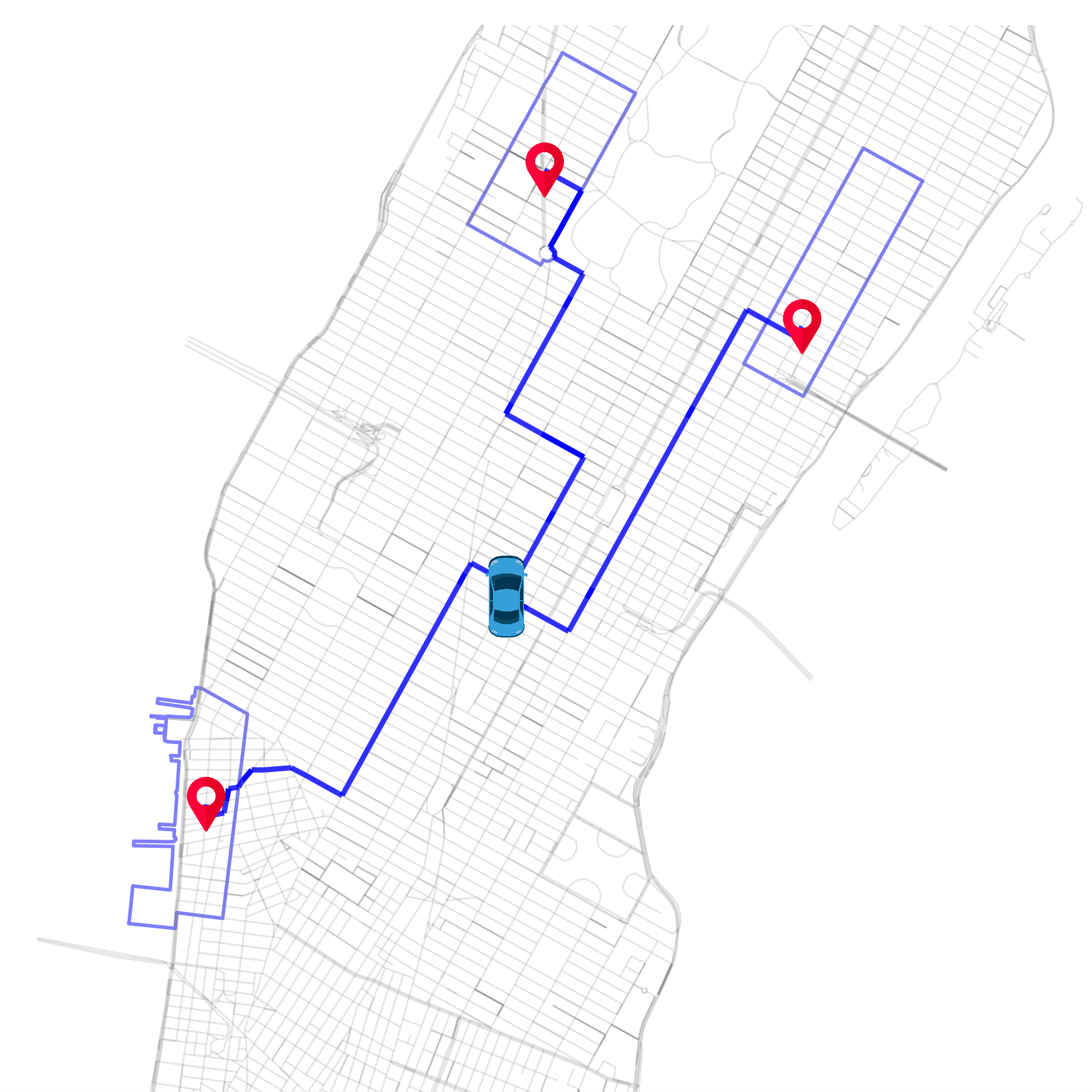}
    \caption{This figure shows a taxi and three potential repositioning regions (with blue boundary). K-means clustering is used to calculate the region centers (Red). Dijkstra’s shortest path algorithm is used to calculate the shortest distance routes between the region centers.}
    \label{fig:driver_choice}
\end{figure}
Then we determine the centers of each region using k-means clustering. We leverage the underlying road network to define a directed graph with the nodes representing the region centers, and the edges representing the shortest path between the region centers. 
We use Dijkstra’s shortest path algorithm on the road network to calculate the shortest distance route between the region centers.
Taxi drivers pick up passengers from one region and drop them off at the other region. After dropping off the customer, a driver has a choice to either keep waiting or reposition to a new region. \cref{fig:driver_choice} shows an idle-standing driver and three potential destinations that will maximize his likelihood of getting the ride. The repositioning recommendation model is triggered at the beginning of each time-step \(k\), i.e., every \(H = 5\) minutes, to ensure rapid repositioning recommendations and avoid scenarios where the taxi driver has to wait for the recommendation.

Driver repositioning preferences are learned by training a separate logistic regression model for each driver based on features including travel distance, expected pickup opportunities, and time of day. Furthermore, XGBoost models are trained to forecast demand in each region and estimate travel times between regions. Detailed descriptions of the training methodology and model performance are provided in Sections~\ref{app:model_training_driver_preference}, \ref{app:model_training_demand}, and \ref{app:model_training_travel_time}.

\subsection{Adherence Aware Vehicle Rebalancing (AAVR) Model Performance}
\label{subsec:aavr-model-performance}

We evaluate the proposed AAVR models in a simulation with 3000 taxis and repeated recommendation decisions over a 5-minute planning horizon. The fleet size is chosen following \cite{wallar2018vehicle}, which showed that approximately 3000 taxis are sufficient to serve most passenger demand in Manhattan when vehicle rebalancing is employed. A 5-minute planning horizon is adopted to enable frequent repositioning recommendations, allowing idle drivers to be redirected promptly in response to changing demand conditions.

Demand, travel time, and driver preferences are forecasted using their respective prediction models discussed above. Each taxi is initialized with a driver confidence value representing the driver's obedience to the recommender system. The initial driver confidence distribution is obtained from the survey-based obedience estimates in \cite{chen2024rebalance}, which are converted into beta priors for each driver. We then perform two case studies: one in which confidence is held fixed by setting the confidence update rates to zero, and one in which confidence evolves according to the beta model. This allows us to separate the allocation benefit of the models from the additional effect of confidence dynamics.

\begin{figure}[H]
\centering
\includegraphics[width=0.5\textwidth]{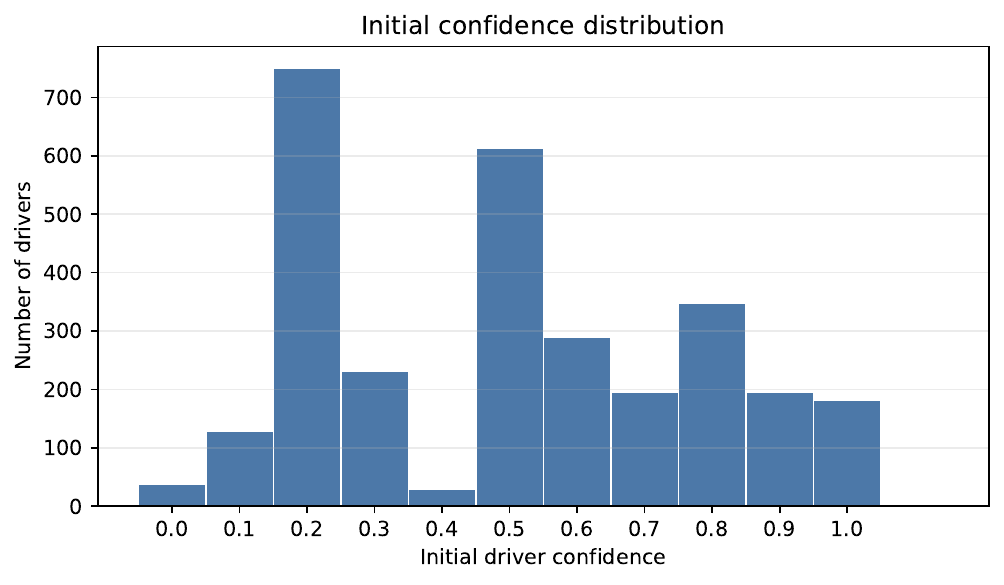}
\caption{Initial driver confidence distribution.}
\label{fig:initial-confidence-distribution}
\end{figure}

Figure~\ref{fig:initial-confidence-distribution} shows the resulting initial fleet confidence distribution used to initialize the driver-specific beta priors.

\subsubsection{Benchmark Methods}

To assess the effectiveness of the proposed framework, we compare AAVR against several benchmark rebalancing strategies. The benchmark methods were selected to represent a broad spectrum of existing repositioning strategies, ranging from no-intervention and heuristic policies to demand-driven optimization methods and adherence-aware methods.

\begin{enumerate}
\item \textbf{Preference (No Recommendation)}: Each driver is assigned to the reachable region, sampling from the driver's learned preference distribution. This is used as the reference policy, since this accounts for no-rebalancing intervention. 

\item \textbf{Random}: Each available driver is assigned uniformly at random among reachable regions. This policy captures an uninformed allocation baseline.

\item \textbf{Hotspot Greedy}: Drivers are assigned greedily to high-demand regions subject to reachability. This baseline prioritizes immediate demand coverage in a decentralized manner.

\item \textbf{Proportional}: Supply is distributed in proportion to region demand following the taxi rebalancing approach in \cite{miao2015taxi}. This encourages demand-weighted coverage across regions.

\item \textbf{Hotspot Aligned}: Recommendations follow the hotspot-aligned rebalancing logic of \cite{wallar2018vehicle}. This model maximizes the product of expected supply and demand in each region while restricting the oversupply to a particular region.

\item \textbf{Matching}: Drivers and regions are matched using the taxi re-positioning method in \cite{lindstroem2025taxi}, which considers driver compliance. The model balances reachable supply with expected region demand.

\item \textbf{AAVR(M)}: This is Problem~\ref{prob:reb_opt} without the variance objective. It maximizes expected served demand while accounting for driver confidence and repositioning preference.

\item \textbf{AAVR}: This is exactly the variance-aware rebalancing optimization model in Problem~\ref{prob:reb_opt}. It penalizes confidence variance in the recommendation objective to improve both allocation performance and steady-state confidence.

\end{enumerate}

The benchmark methods differ substantially in how they incorporate demand information, driver preferences, and adherence uncertainty. While some methods focus solely on balancing supply and demand, others explicitly account for driver behavior when generating recommendations. AAVR(M) uses driver confidence and preference to model expected adherence, whereas AAVR additionally incorporates adherence variability through the variance-aware objective in Problem~\ref{prob:reb_opt}. To quantify the benefits of these design choices, we evaluate each method using a set of operational, economic, and behavioral performance metrics.

\subsubsection{Evaluation Metrics}

The following metrics are used to evaluate each method:

\begin{enumerate}
\item \textbf{Allocated requests}: Total number of passenger requests that are successfully served by the recommended taxi allocation. This measures the demand-coverage performance of each policy.

\item \textbf{Platform earnings}: Total revenue earned by the platform from served passenger requests. This captures the operator-side economic value generated by each recommendation policy.

\item \textbf{Driver earnings}: Average driver profit after accounting for both relocation travel cost and service travel earnings. This metric reflects whether the policy improves driver-side economic outcomes in addition to platform performance.

\item \textbf{Passenger waiting time}: Average time a passenger waits before being allocated to a taxi. Passengers who are not allocated by the end of the 5-minute planning horizon are assumed to drop out, so lower values indicate faster passenger service.

\item \textbf{Fleet confidence}: Average confidence of the taxi fleet before recommendations are issued. This metric captures how the recommendation policy affects driver willingness to follow future rebalancing instructions.

\end{enumerate}

The tables report the performance of each method relative to the Preference benchmark, with positive values indicating improvements over Preference and negative values indicating worse performance. For passenger waiting time, however, negative values correspond to shorter waiting times and are therefore desirable. The right panels of the figures report the corresponding percentage improvements over the best non-AAVR benchmark.

\subsubsection{Case Study I: Fixed Driver Confidence}

We first isolate the effect of the rebalancing policy by disabling confidence evolution. Specifically, the confidence update rates are set to zero, causing the confidence distribution to remain fixed throughout the simulation horizon. Under this setting, adherence probabilities do not evolve over time, and any performance differences can be attributed solely to the recommendation and allocation logic of the competing methods.

\begin{table}[tbh]
\centering
\caption{Performance relative to the Preference benchmark under fixed driver confidence. Entries report $(\mathrm{Method}-\mathrm{Preference})$. Positive values indicate an improvement over Preference, except for passenger waiting time, where negative values are preferable. Bold entries indicate the best value in each row.}
\label{tab:aavr-no-confidence-change}
\resizebox{\textwidth}{!}{%
\begin{tabular}{lrrrrrrr}
\toprule
Metric & Random & Hotspot Greedy & Proportional & Hotspot Aligned & Matching & AAVR(M) & AAVR \\
\midrule
Allocated requests & -7,674 & -33,615 & 106,192 & 148,253 & 107,980 & 183,246 & \textbf{188,977} \\
Platform earnings & -26,936 & -162,305 & 481,082 & 654,915 & 473,271 & 805,764 & \textbf{831,615} \\
Driver earnings & -0.156 & -0.525 & 1.256 & 1.683 & 1.246 & 2.087 & \textbf{2.181} \\
Passenger waiting time & 0.038 & 0.155 & -0.505 & -0.707 & -0.514 & -0.874 & \textbf{-0.901} \\
Fleet confidence & 0.000 & 0.000 & 0.000 & 0.000 & 0.000 & 0.000 & 0.000 \\
\bottomrule
\end{tabular}%
}
\end{table}

\begin{figure}[tbh]
\centering
\includegraphics[width=\textwidth]{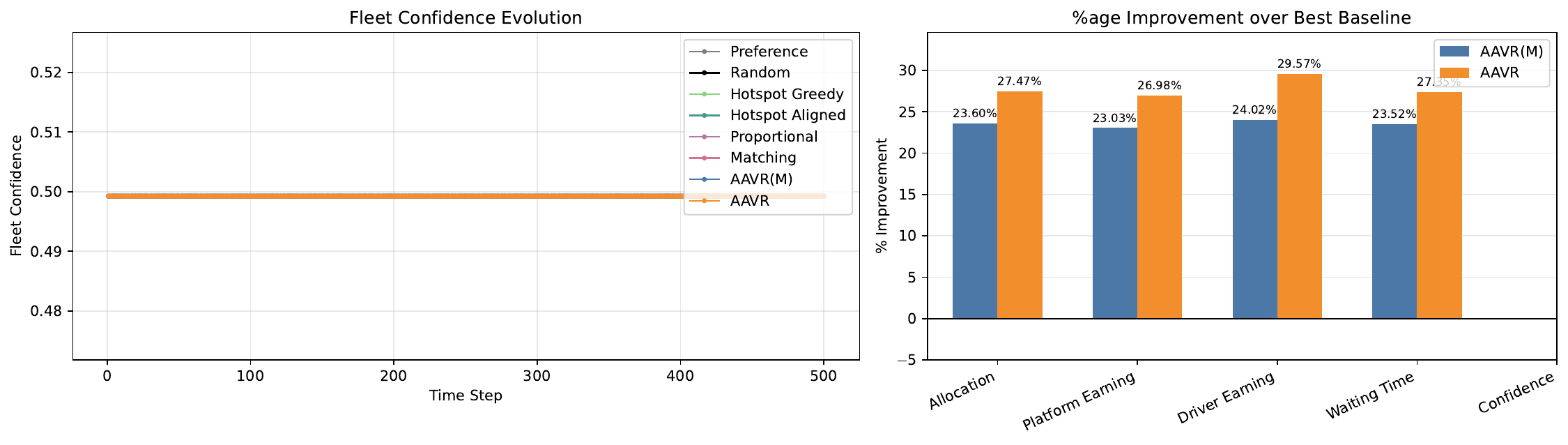}
\caption{Comparison of benchmark methods under fixed driver confidence. Left: fleet confidence evolution. Right: percentage improvement over the best non-AAVR baseline.}
\label{fig:aavr-no-confidence-change}
\end{figure}

Table~\ref{tab:aavr-no-confidence-change} shows that the Random and Hotspot Greedy baselines perform worse than the Preference benchmark across most metrics, highlighting that no recommendation is better than random or a greedy recommendation strategy. Among the benchmark methods, Hotspot Aligned achieves the best overall performance.

Figure~\ref{fig:aavr-no-confidence-change} reports the corresponding percentage improvement over this best baseline. With fixed confidence, AAVR improves over Hotspot Aligned by 27.47\% in allocation, 26.98\% in platform earnings, 29.57\% in driver earnings, and 27.35\% in passenger waiting time. AAVR(M) also improves over Hotspot Aligned, but by a smaller margin across all four metrics. 
These results indicate that incorporating the variance-aware objective provides additional benefits beyond maximizing expected served demand alone. 

This additional improvement is consistent with the theoretical analysis in \cref{sec:theoretical}. While AAVR(M) aligns the expected supply with the expected demand through the mean-matching objective, AAVR additionally minimizes the variance of the realized supply. As established by the tractable reformulation, reducing the variance of the realized supply yields a tighter approximation to the original cost-of-deviation objective, resulting in more reliable supply-demand matching and improved operational performance. Since confidence updates are disabled, all methods have identical fleet confidence values, as expected.

\subsubsection{Case Study II: Evolving Driver Confidence}

We next consider a setting in which driver confidence evolves according to the Beta opinion update model. The standard Beta--Bernoulli update, as in~\cite{josang2002beta}, uses $\epsilon_1(c)=\epsilon_0(c)=1$, so that each successful or unsuccessful outcome contributes one unit of evidence to the confidence estimate. However, driver sensitivities need not be identical across the fleet. Moreover, a driver's sensitivity to a successful outcome, i.e., an allocated recommendation, may differ from their sensitivity to a failed outcome, i.e., a recommendation that does not result in allocation. Therefore, instead of keeping these values fixed, we allow them to be heterogeneous across drivers. For each driver $c$, the positive- and negative-evidence sensitivities are drawn independently as
\[
\epsilon_1(c) \sim \mathrm{Uniform}(0.1,1.9),
\qquad
\epsilon_0(c) \sim \mathrm{Uniform}(0.1,1.9),
\]
and then held fixed over time. Thus, drivers differ not only in their initial confidence, but also in how strongly they respond to successful and unsuccessful recommendation outcomes. Consequently, recommendation quality affects both immediate allocation outcomes and future adherence behavior through heterogeneous confidence adaptation.

\begin{table}[tbh]
\centering
\caption{Performance relative to the Preference benchmark under evolving driver confidence. Entries report $(\mathrm{Method}-\mathrm{Preference})$. Positive values indicate an improvement over Preference, except for passenger waiting time, where negative values are preferable. Bold entries indicate the best value in each row.}
\label{tab:aavr-confidence-change}
\resizebox{\textwidth}{!}{%
\begin{tabular}{lrrrrrrr}
\toprule
Metric & Random & Hotspot Greedy & Proportional & Hotspot Aligned & Matching & AAVR(M) & AAVR \\
\midrule
Allocated requests & -6,481 & 2,073 & 116,662 & 160,148 & 130,959 & 194,316 & \textbf{202,947} \\
Platform earnings & -22,804 & 163 & 526,231 & 704,580 & 572,012 & 848,626 & \textbf{887,080} \\
Driver earnings & -0.129 & -0.077 & 1.371 & 1.806 & 1.506 & 2.190 & \textbf{2.325} \\
Passenger waiting time & 0.033 & -0.016 & -0.555 & -0.766 & -0.625 & -0.928 & \textbf{-0.969} \\
Fleet confidence & -0.127 & -0.109 & 0.075 & 0.042 & 0.063 & 0.078 & \textbf{0.098} \\
\bottomrule
\end{tabular}%
}
\end{table}

\begin{figure}[tbh]
\centering
\includegraphics[width=\textwidth]{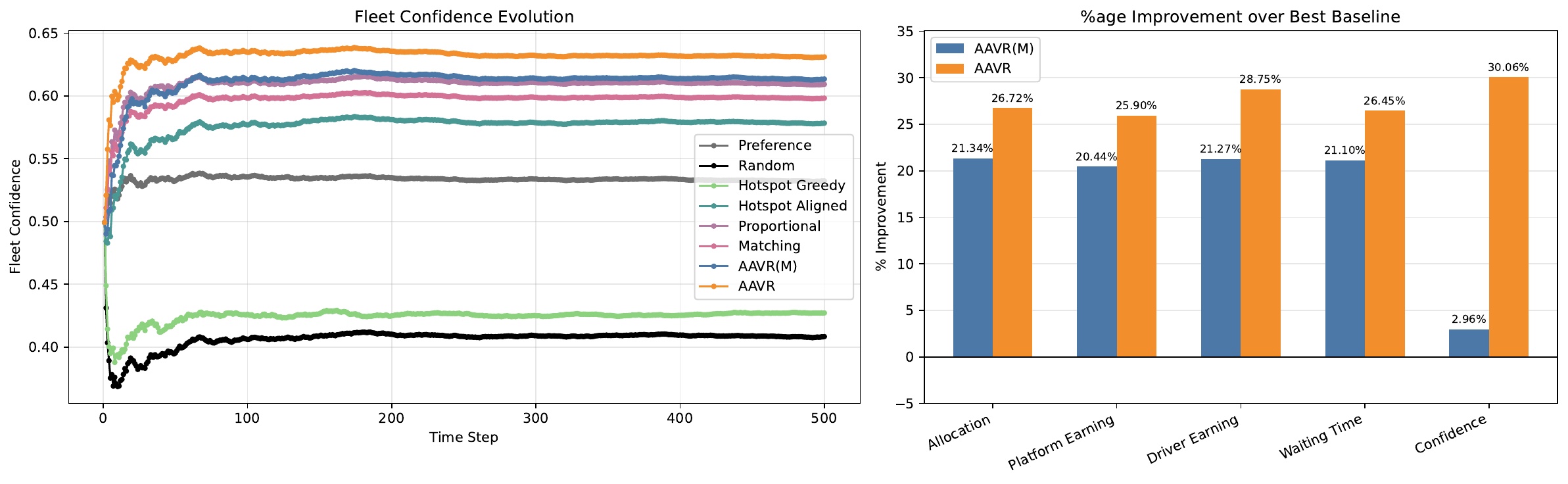}
\caption{Comparison of benchmark methods under evolving driver confidence. Left: fleet confidence evolution. Right: percentage improvement over the best non-AAVR baseline.}
\label{fig:aavr-confidence-change}
\end{figure}

The results are reported in Table~\ref{tab:aavr-confidence-change}. As confidence evolves over time, recommendation quality influences both the immediate allocation outcome and future driver adherence. Policies that consistently generate successful recommendations maintain or increase driver confidence, leading to higher adherence in subsequent planning horizons. Conversely, unsuccessful recommendations reduce driver confidence, diminishing future adherence and ultimately degrading operational performance. Because $\epsilon_1(c)$ and $\epsilon_0(c)$ are driver-specific and fixed over time, the strength of this behavioral feedback differs across drivers, creating a heterogeneous closed-loop interaction between recommendation quality, driver confidence, and system performance.

Figure~\ref{fig:aavr-confidence-change} summarizes the percentage gains over the best non-AAVR baseline. For allocation, platform earnings, driver earnings, and passenger waiting time, Hotspot Aligned is again the strongest baseline. AAVR improves over Hotspot Aligned by 26.72\% in allocation, 25.90\% in platform earnings, 28.75\% in driver earnings, and 26.45\% in passenger waiting time. For fleet confidence, Proportional is the strongest baseline, and AAVR improves over it by 30.06\%. In comparison, AAVR(M) achieves smaller improvements of 21.34\%, 20.44\%, 21.27\%, 21.10\%, and 2.96\% for allocation, platform earnings, driver earnings, passenger waiting time, and fleet confidence, respectively. This indicates that modeling expected adherence alone is helpful, but less effective than also accounting for adherence variability.

Figure~\ref{fig:aavr-confidence-change} also shows the evolution of fleet confidence over time. Unlike the fixed-confidence setting, confidence trajectories now evolve as drivers accumulate recommendation experiences. AAVR achieves the highest steady-state confidence levels while simultaneously achieving the strongest operational performance. This suggests that variance-aware recommendations create a positive feedback mechanism in which successful recommendations increase confidence, higher confidence improves future adherence, and improved adherence further enhances system performance.

Overall, the two case studies demonstrate that modeling adherence uncertainty improves both immediate operational performance and long-term fleet behavior. When confidence dynamics are included, the proposed AAVR framework simultaneously improves allocation efficiency, platform and driver earnings, passenger waiting time, and driver confidence.

\section{Conclusion}
\label{sec:conc}

This paper introduced the Adherence-Aware Vehicle Rebalancing (AAVR) framework, a human-centered approach to taxi rebalancing that explicitly accounts for the uncertainty induced by driver adherence to repositioning recommendations. Unlike conventional rebalancing models that assume recommendations are followed deterministically, the proposed framework recognizes that realized supply is ultimately generated through driver decisions. In practice, a driver's willingness to follow a recommendation depends not only on the driver's intrinsic destination preferences but also on the driver's confidence in the recommender system. Moreover, this confidence evolves over time through repeated interactions with the platform, implying that recommendation policies influence both immediate operational outcomes and future driver behavior.

To capture these dynamics, we developed a behavioral model in which adherence is jointly determined by destination preference and dynamically evolving confidence. Driver confidence was represented using a Beta confidence model whose updates satisfy natural behavioral consistency properties, ensuring that successful recommendations increase confidence while unsuccessful recommendations reduce it. These confidence estimates were then incorporated into the recommendation-generation framework.

Building on this behavioral model, we formulated the Adherence-Aware Vehicle Rebalancing optimization problem. Since recommendations only influence the probability of driver repositioning, the resulting realized supply is stochastic, leading to a computationally intractable optimization problem. To address this challenge, we derived an upper bound on the original objective and used it to construct an adherence-aware optimization model suitable for real-time recommendation generation in large ride-hailing fleets.

The proposed framework was evaluated using a simulation environment calibrated with New York City taxi data. The experiments considered both static-confidence and dynamic-confidence settings and compared the proposed method against multiple state-of-the-art rebalancing benchmarks. Across all evaluated scenarios, AAVR consistently achieved superior performance in terms of passenger allocation, platform earnings, driver earnings, and passenger waiting time. Furthermore, when confidence dynamics were enabled, the proposed framework produced the largest improvements in fleet confidence, demonstrating its ability to improve both short-term operational efficiency and long-term driver trust. The results further showed that explicitly accounting for adherence variability yields additional benefits beyond considering expected adherence alone.

Several directions remain for future research. In particular, the proposed framework can be extended to jointly optimize rebalancing recommendations and incentive design while accounting for the social dynamics of driver confidence. Rather than updating confidence solely through individual experiences, future models could incorporate opinion formation over driver social networks, allowing confidence and adherence behavior to evolve through both direct recommendation outcomes and peer influence. Such a framework would enable the study of how incentives and information propagation jointly shape long-term fleet-wide adherence and system performance.

\section*{Acknowledgment}

This work was conducted independently and has not been sponsored, reviewed, or endorsed by Uber. The views, findings, and conclusions expressed in this paper are solely those of the authors and do not necessarily reflect those of Uber.

 \bibliographystyle{elsarticle-num-names}

\bibliography{bibsample}

\appendix
\section{Additional Computational Results}
\label{app:additional-computational-results}

\subsection{Convex-Order Optimality at Fixed Expected Supply}
\label{app:convex-order}
To emphasize the dependence of the supply random variable on the decision vector, we write $\mathbf{S}(x)$ in this section.

\begin{theorem}[Optimality under convex-order minimality]
\label{thm:convex-order-optimality}
Let
\begin{equation}
\mathcal{X}_{m^\star}
=
\{x:\mathbb{E}[\mathbf{S}(x)]=m^\star\},
\end{equation}
and assume that $\mathbf{D}$ is independent of $\mathbf{S}(x)$ for every
$x\in\mathcal{X}_{m^\star}$. If there exists
$x^\dagger\in\mathcal{X}_{m^\star}$ such that
\begin{equation}
\mathbf{S}(x^\dagger)\le_{cx}\mathbf{S}(x)
\quad \text{for all } x\in\mathcal{X}_{m^\star},
\end{equation}
then $x^\dagger$ minimizes $\mathcal{J}(x)$ over
$\mathcal{X}_{m^\star}$.
\end{theorem}

\begin{proof}
Using $g(s)=\mathbb{E}_{\mathbf{D}}[|\mathbf{D}-s|]$, since $s\mapsto |d-s|$ is convex for every fixed $d$, it follows that
$g$ is convex. Therefore,
\begin{equation}
\mathbf{S}(x^\dagger)\le_{cx}\mathbf{S}(x)
\quad \Longrightarrow \quad
\mathbb{E}[g(\mathbf{S}(x^\dagger))]
\le
\mathbb{E}[g(\mathbf{S}(x))].
\end{equation}
Moreover, by the independence of $\mathbf D$ and $\mathbf S(x)$,
\[
\mathbb E[g(\mathbf S(x))]
=
\mathbb E_{\mathbf D,\mathbf S(x)}
\!\left[|\mathbf D-\mathbf S(x)|\right]
=
\mathcal J(x),
\]
and
$\mathbb E[g(\mathbf S(x^\dagger))]=\mathcal J(x^\dagger)$.
Therefore, $\mathcal J(x^\dagger)\le \mathcal J(x)$ for all
$x\in\mathcal X_{m^\star}$, and hence $x^\dagger$ minimizes
$\mathcal J(x)$ over $\mathcal X_{m^\star}$.
\end{proof}

\begin{remark}[Role of box constraints]
Convex-order minimality gives an exact sufficient condition for optimality at fixed expected supply. By \cref{lem:maj-to-cx}, for Poisson--binomial supplies, this condition is satisfied if the corresponding probability vector majorizes all other feasible probability vectors. However, heterogeneous box constraints \(\ell_c\le p_c(x)\le u_c\) may prevent the existence of such a globally majorizing vector. The following example illustrates this limitation.
\end{remark}

\begin{example}[Non-existence of a globally majorizing vector]
\label{ex:non-existence-majorization}
Consider \(n=3\) drivers. Let
\(p=(p_1,p_2,p_3)\), where \(p_i\) denotes the arrival probability of driver \(i\) as defined in \eqref{eqn:arrival_prob_ss}. Suppose the expected target supply is \(p_1+p_2+p_3=1.32\), with bounds \(p_1\in[0.39,0.46]\),\(p_2\in[0.36,0.51]\), and \(p_3\in[0.35,0.57]\).

Consider the feasible vector
\(u=(0.39,0.36,0.57)\), with
\(u^\downarrow=(0.57,0.39,0.36)\).
Since \(0.57\) is the largest attainable component value under the box constraints and the total sum is fixed at \(1.32\), any feasible vector that majorizes \(u\) must coincide with \(u\). Thus, if a feasible vector majorizing all feasible vectors exists, it must be \(u\).

Now consider the feasible vector
\(v=(0.46,0.51,0.35)\), with
\(v^\downarrow=(0.51,0.46,0.35)\).
Since
\(0.57+0.39 < 0.51+0.46\),
we have \(u \not\succeq v\).
Hence, no feasible vector can majorize every feasible vector, demonstrating that a globally majorizing vector need not exist.
\end{example}

\subsection{Driver Preference Prediction}
\label{app:model_training_driver_preference}
\begin{figure}[H]
    \centering
    \includegraphics[width=0.7\linewidth]{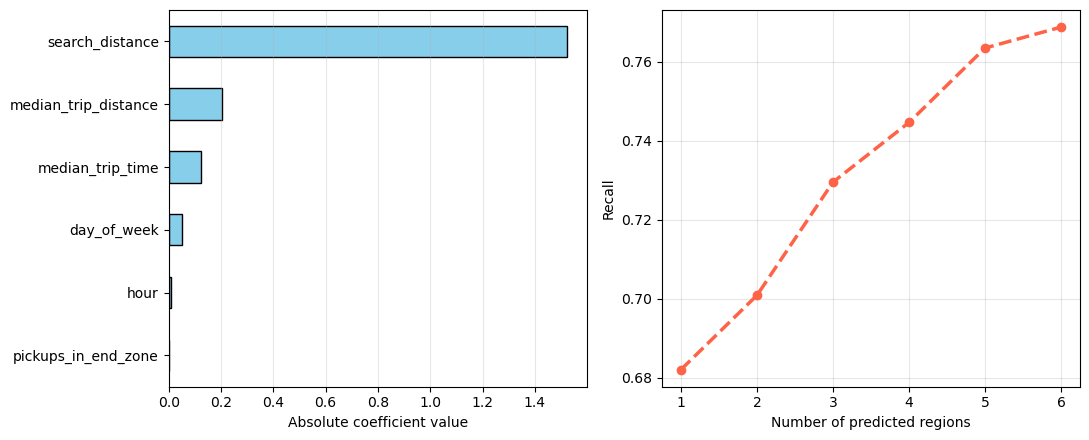}
    \caption{Feature importance and top-$k$ accuracy for the driver preference prediction model.}
    \label{fig:preference_prediction}
\end{figure}

The driver preference prediction model is trained using historical driver trip records from \cite{donovan2014new}. For each observed decision, the driver's next destination region is treated as the chosen alternative, while other candidate regions are treated as non-chosen alternatives. The model uses features describing each candidate region: search distance from the driver's current region, expected pickups in the candidate region, median trip time and median trip distance of trips originating from the candidate region, hour of day, and day of week. We evaluate the model using top-$k$ accuracy. For each decision instance, the model selects the $k$ regions with the highest predicted preference scores. If the driver's actual destination is among these $k$ regions, the prediction is counted as correct. Top-$k$ accuracy is the fraction of correct predictions over all decision instances. As shown in the left plot of \cref{fig:preference_prediction}, search distance is the dominant feature, followed by median trip distance and median trip time. Temporal features have smaller effects, while expected pickups in the candidate region has little influence. The right plot shows that top-$k$ accuracy increases from approximately 68.2\% for $k=1$ to 70.1\%, 72.9\%, 74.5\%, 76.4\%, and 76.9\% for $k=2,\ldots,6$, respectively. This indicates that the model assigns high preference scores to regions drivers are likely to choose. For repositioning, the objective is not necessarily to predict a single destination, but to estimate a probability distribution over plausible destination regions. These probabilities are subsequently used as the preference probabilities \(L_{cj}(k)\) in the AAVR model to estimate the realized supply distribution.

\subsection{Demand Prediction}
\label{app:model_training_demand}

Demand prediction was formulated as a time-series forecasting problem. An XGBoost model~\citep{chen2016xgboost} was trained using the features described in \cref{sec:demand-travel-time}. The look-back window, number of estimators, maximum tree depth, and learning rate were optimized through grid search. To benchmark the proposed demand prediction pipeline, we additionally compare XGBoost with two state-of-the-art pre-trained time-series foundation models, CHRONOS~\citep{ansari2024chronos} and MOIRAI~\citep{woo2024unified}.

\begin{figure}[tbh]
    \centering
    \includegraphics[width=0.7\textwidth]{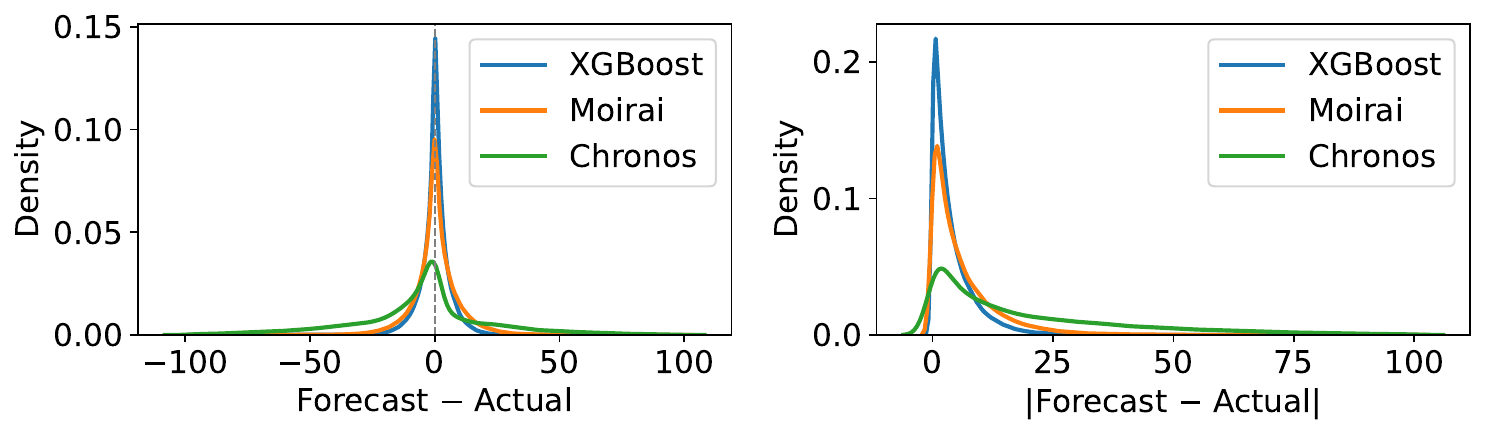}
    \caption{Comparison of demand prediction errors. Left: error distribution. Right: absolute error distribution.}
    \label{fig:demand forecasting}
\end{figure}

Table~\ref{tab:error_statistics} shows that XGBoost achieved the best performance, with the lowest MAE (4.05), RMSE (6.26), and percentile errors, outperforming both MOIRAI and CHRONOS. Figure~\ref{fig:demand forecasting} further compares the error distributions of the three models. 

\begin{table}[tbh]
\centering
\renewcommand{\arraystretch}{0.9}
\caption{Error statistics of demand prediction models.}
\begin{tabular}{lccc}
\toprule
\textbf{Metric} & \textbf{XGBoost} & \textbf{MOIRAI} & \textbf{CHRONOS} \\
\midrule
Mean Absolute Error & 4.05 & 6.47 & 29.99 \\
Root Mean Squared Error & 6.26 & 13.06 & 106.96 \\
25th Percentile Error & 0.90 & 1.32 & 3.81 \\
75th Percentile Error & 5.54 & 8.58 & 32.00 \\
95th Percentile Error & 13.23 & 20.74 & 77.00 \\
\bottomrule
\end{tabular}
\renewcommand{\arraystretch}{1}
\label{tab:error_statistics}
\end{table}

\subsection{Travel Time Prediction}
\label{app:model_training_travel_time}

\begin{figure}[tbh]
    \centering
    \includegraphics[width=0.7\textwidth]{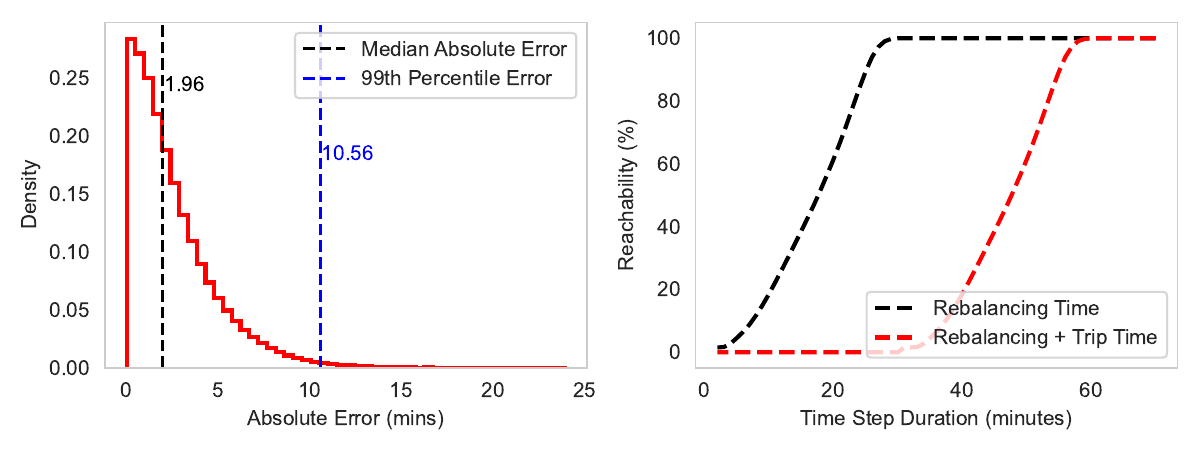}
    \caption{Left: distribution of absolute travel time prediction errors. Right: expected reachability as a function of the planning horizon.}
    \label{fig:travel time error}
\end{figure}

Figure~\ref{fig:travel time error} summarizes the travel time prediction performance. The median absolute prediction error is approximately 2 minutes, while the 99th percentile error is about 11 minutes. Predicted expected travel times determine the expected reachability of idle taxis, defined as the fraction of origin--destination region pairs whose predicted travel time is within the planning horizon~\citep{brar2021dynamic}. As shown in Figure~\ref{fig:travel time error}, a planning horizon of approximately 30 minutes enables expected repositioning between nearly all Manhattan regions, while a 60-minute horizon is sufficient to accommodate nearly all inter-region repositioning and passenger trips. In our implementation, we use a 5-minute planning horizon to provide frequent repositioning recommendations. Although this reduces the set of expectedly reachable destinations within a single planning horizon, it aligns with drivers' preference for shorter repositioning trips and enables rapid online updates.

\end{document}